\documentclass[aps,prd,twocolumn,showpacs,superscriptaddress,groupedaddress]{revtex4} 
\usepackage{mathrsfs}
\usepackage{epsfig}
\usepackage{graphicx}
\usepackage{mathtools}
\usepackage{color}
\usepackage{amssymb}

\newcommand{\HI}{\rm{21\,cm}}
\newcommand{\Ha}{\rm{H}\alpha}
\newcommand{\ly}{\rm{Ly}\alpha}

\begin{document}
\title{A Halo Model Approach to the $\HI$ and $\ly$ Cross-correlation}

\author{Chang Feng\footnote{chang.feng@uci.edu}}
\affiliation{Department of Physics and Astronomy,
University of California, Irvine, CA 92697, USA }

\author{Asantha Cooray} 
\affiliation{Department of Physics and Astronomy,
University of California, Irvine, CA 92697, USA }

\author{Brian Keating} 
\affiliation{Department of Physics, University of California, San Diego, CA 92093, USA}

\begin{abstract}
We present a halo-model-based approach to calculate the cross-correlation between $\HI$ HI intensity fluctuations 
and $\ly$ emitters (LAE) during the epoch of reionization (EoR). Ionizing radiation around dark matter halos are modeled as bubbles with the size and growth determined based on the reionization photon production, among
other physical parameters. The cross-correlation shows a clear negative-to-positive transition, associated with transition from ionized to neutral hydrogen in the intergalactic medium during EoR. 
The cross-correlation is subject to several foreground contaminants, including foreground radio point sources important for $\HI$ experiments and low-$z$ interloper emission lines, such as $\Ha$, OIII, and OII, for $\ly$ experiments. 
Our calculations show that by masking out high fluxes in the $\ly$ measurement, the correlated foreground contamination on the $\HI$\mbox{--}$\ly$ cross-correlation can be dramatically reduced. 
We forecast the detectability of $\HI$\mbox{--}$\ly$ cross-correlation at different redshifts and adopt a Fisher matrix approach to estimate uncertainties on the key EoR parameters that have not 
been well constrained by other observations of reionization. This halo-model-based approach enables us to  explore the EoR parameter space rapidly for different $\HI$ and $\ly$ experiments.
\end{abstract}

\maketitle

\section{Introduction}

The early universe, initially filled with hot plasma, became neutral as hydrogen ions captured electrons that were decoupled from cosmic microwave background (CMB) photons at a redshift of 1100. 
A cosmic ``dark age'' subsequently ensued in the universe until the linear density fluctuations seeded by inflation were amplified, forming the first stars and galaxies~\cite{eorreview}. The X-rays from mini quasars and ultraviolet radiation from the massive stars in first-light galaxies heated and ionized the neutral hydrogen, and the universe gradually transformed from completely neutral to fully ionized during the epoch of reionization (EoR). Today the EoR still remains largely unexplored, as signatures imprinted on the intergalactic medium (IGM) in the early universe are too faint to be detected.

The physical processes present during the EoR are of extreme importance to our understanding of the universe and the structure that formed in it.
The clustering of neutral hydrogen (HI) down to the Jeans length scale contains a wealth of information about certain fundamental physics, including dark matter. 
The HI tomography is not subject to small-scale physical effects such as photon diffusion damping present in the CMB power spectrum. 
The timing and duration of the EoR can help interpret other cosmological measurements, such as the kinetic Sunyaev-Zel'dovich (kSZ) effect~\cite{2005ApJ...630..643M}. 
Moreover, some exotic physics such as primordial magnetic fields~\cite{2009ApJ...692..236S} and decaying dark matter~\cite{2006PhRvD..74j3502F} could be probed during the EoR. 
To date, the neutral fraction during the EoR was measured from quasar absorption spectra~\cite{2006ARA&A..44..415F} and $\ly$-emitting galaxy luminosity functions~\cite{2014ApJ...797...16K,2004ApJ...617L...5M, 2006ApJ...647L..95M} around $z\sim6$. Another important quantity of the EoR, the Thomson scattering optical depth, is constrained to $\tau$ = 0.088 $\pm$ 0.014 by WMAP~\cite{wmap7yr} and $\tau$ = 0.058 $\pm$ 0.012 by Planck satellites~\cite{plancktau2016}.

The best way to measure the HI content prior to and during reionization is through the $\HI$ HI fine-structure spin-flip transition. 
A number of experiments have been targeting the $\HI$ emission, such as the Low Frequency Array (LOFAR)~\cite{2013A&A...556A...2V}, the Murchison Widefield Array (MWA)~\cite{2013PASA...30....7T}, the Precision Array for Probing the Epoch of Reionization (PAPER)~\cite{2010AJ....139.1468P}, the Hydrogen Epoch of Reionization Array (HERA)~\cite{2016arXiv160607473D} and the Square Kilometer Array (SKA)~\cite{2015aska.confE...1K}. The redshifted $\HI$ emission is contaminated by both galactic and extragalactic foregrounds that consist of galactic synchrotron, supernovae remnants, free-free emission, and radio point sources~\cite{2006PhR...433..181F}. The Galactic synchrotron emission is the dominant contribution, as it is three to four orders of magnitude stronger than the background brightness temperature fluctuations. By performing a component separation or subtracting the $\HI$ foreground, the $\HI$ brightness fluctuations could be measured~\cite{2015MNRAS.447.1973B}. This, however, relies on the reliability of the foreground estimation. The radio point sources are also thought to be another foreground issue for $\HI$ experiments, but this signal 
is very likely to be a subdominant contamination~\cite{2009MNRAS.394.1575L}. The expected $\HI$ signal is at the level of 10 $\rm{mK}^2$ at $k=0.3$ $\rm{Mpc}^{-1}$~\cite{2011MNRAS.411..955M}, while
recent measurements from PAPER set a 2$\sigma$ upper limit as ($22.4\, \rm{mK})^2$ in the range $0.15<k<0.5$ $h\,\rm{Mpc}^{-1}$ at $z=8.4$~\cite{2015ApJ...809...61A}.

During the EoR the ultraviolet $\ly$ emission was created by the first stars and galaxies. The $\ly$ background traces the underlying dark matter distribution and also affects the spin-temperature distribution. By directly measuring the $\ly$ emissions, we get an additional observable on EoR physics as well~\cite{2013MNRAS.428.1366J}. However, the $\ly$ background is contaminated by low-$z$ foregrounds, such as $\Ha$ at $z$ = 0.5, OIII at $z$ = 0.9, and OII at $z$ = 1.6. These low-$z$ components are much brighter than $\ly$, precluding a clean detection. On the other hand, such low-$z$ foregrounds can be easily masked out since they are very bright~\cite{pullen13,yanlya}. 
Therefore, a simple masking procedure would recover the genuine $\ly$ background from experiments. 

The $\HI$ and $\ly$ emission is anti-correlated at large angular scales because they originate from IGM and galaxies, respectively, and ionized bubbles around $\ly$ galaxies are devoid of HI that is seen with $\HI$ experiments. 
The transition in the cross-correlation from negative to positive indicates a characteristic size for the average of HII regions around halos. 
Therefore, the cross-correlation between $\HI$ and $\ly$ can be viewed as a complementary probe of EoR physics. The cross-correlation could be more advantageous in terms of foreground removal as the two sets of aforementioned foregrounds would be largely uncorrelated, potentially allowing a higher signal-to-noise detection and an easy confirmation of the EoR signature. Previously, the cross-correlation between $\HI$ experiments and galaxies was studied for $\HI$ experiments such as MWA and LOFAR, using both analytical and numerical calculations~\cite{2007ApJ...660.1030F,lidz21cm,2007MNRAS.375.1034W}, as well as for LOFAR and Subaru's Hyper Suprime-cam~\cite{2016MNRAS.457..666V}. The cross-correlation between $\HI$ and CO/kSZ also shows a similar transition in the correlation sign~\cite{2011ApJ...741...70L,2010MNRAS.402.2279J}.

So far, different approaches have been used to model reionization. The large-scale $N$-body and radiative transfer simulations, while desirable, are challenging because it is computationally intensive due to the large dynamic range~\cite{2006MNRAS.369.1625I,Iliev}. Another approach involves semi-analytical/semi-numerical models by taking a halo catalog generated from $N$-body simulations and generating a reionization field by smoothly filtering the halo field~\cite{2007ApJ...654...12Z, 2011MNRAS.414..727Z}. A more simplified idea of this semi-numerical simulation is to make the density field from Gaussian random variables instead of relying on the $N$-body simulations. The simulation can be done efficiently within a small box for EoR~\cite{2007ApJ...669..663M,2011MNRAS.411..955M}. However, this numerical solution becomes ineffective when the box is too large or the simulated epoch is far beyond the EoR when the CMB temperature $T_{\rm{cmb}}$ is coupled to the spin temperature $T_{\rm{s}}$ and the assumption $T_{\rm{s}}\gg T_{\rm{cmb}}$ breaks down. An upgraded version of this implementation uses a very similar algorithm to extend to large boxes~\cite{2010MNRAS.406.2421S}. 
Here, we apply a very simple ionizing bubble model~\cite{2004ApJ...613....1F} to the calculations of $\HI$ brightness temperature anisotropy and its cross-correlation with $\ly$ analytically, so
we can quickly forecast the detectability of the signal for different combinations of $\HI$ and $\ly$ experiments, and explore the EoR parameter space without significant computational cost. 
This approach would be very beneficial when the cross-correlation measurements with different experiments and major foreground or instrumental issues need to be identified in the early stage of the development. 

This paper is organized as follows. In Section \ref{model}, we introduce the halo model for the ionizing bubble as well as the cross-correlation. In Section \ref{lya}, the $\ly$ luminosity is discussed. Then we focus on the low-$z$ foregrounds for both $\HI$ and $\ly$ measurements in Section \ref{lowzfg} and estimate signal-to-noise for the detectability of different experiments, as well as the uncertainties on the EoR parameters in Section \ref{forecast}. We conclude in Section \ref{con}. We use the Planck cosmological parameters: $\Omega_bh^2=0.02230$, $\Omega_ch^2=0.1188$, $H_0=67.74\,\rm{km}/\rm{s}/\rm{Mpc}$, $Y_{\rm p}=0.249$, $\ln(10^{10}A_s)=3.064$ at $k_{\ast}=0.05\,\rm{Mpc}^{-1}$, $n_s=0.9667$, and $\tau=0.058$.

\section{Theoretical Model of the Cross-correlation}
\label{model}

Here we describe the basic ingredients of our halo model. Since the mean ionizing fraction is not precisely constrained by current observations, we use
the CAMB's reionization model~\cite{2008PhRvD..78b3002L}; i.e.,
\begin{equation}
\bar x_e(z)=\frac{1}{2}\Big[1+\tanh\Big(\frac{(1+z_{\rm{re}})^{3/2}-(1+z)^{3/2}}{\Delta y}\Big)\Big],
\end{equation}
where the redshift $z_{\rm{re}}$ is derived from the optical depth $\tau$ today; i.e., 
\begin{equation}
\tau=\int^{\chi_{\rm{re}}}_0d\ell n_e(\chi')\sigma_T \, ,
\end{equation}
and $\Delta y$ = $1.5\sqrt{1+z_{\rm{re}}}$$\Delta_z$. Here, $\sigma_T$ is the Thomson cross-section, the electron density is $n_e=(1-3/4Y_{\rm p})\rho_{b,0}/m_{\rm{H}}a^{-3}x_e$, the comoving length $d\ell=ad\chi'$, the Helium fraction is $Y_{\rm p}$, the proton mass is $m_{\rm{H}}$, and the mean neutral hydrogen fraction is $\bar x_H=1-\bar x_e$. We assume that helium is singly ionized along with hydrogen, while the double ionization of helium is neglected.

The $\HI$ brightness temperature can be split into two components, $T_0(z)\psi(\textbf{x},z)$, in which the isotropic background temperature is
\begin{equation}T_0(z)=27\Big(\frac{1-Y_{\rm p}}{1-0.248}\Big)\Big(\frac{\Omega_b}{0.044}\Big)\Big[\Big(\frac{0.27}{\Omega_m}\Big)\Big(\frac{1+z}{10}\Big)\Big]^{1/2}\,\,(\rm{mK}),
\end{equation}
and spatial fluctuation $\psi$ is~\cite{cora2}
\begin{equation}\psi(\textbf{x},z)=\bar x_H(1+\delta_x)(1+\delta)=\bar x_H(1+\delta_x+\delta+\delta_x\delta).\label{com21}
\end{equation}
Here $\delta_x$ is the density contrast of ionizing field ($x$) and we neglect perturbations introduced by spin-temperature fluctuations and peculiar velocities. For the $\HI$ field, we only consider the signals from IGM as galaxy contributions are $10^{-4}$ times smaller~\cite{GongHI-CO}, and model the ionizing field with ``bubbles"~\cite{cora2}. From Eq. (\ref{com21}), the two-point correlation functions for ionizing and matter density contrasts are $\langle\delta_x\delta_x\rangle=\xi_{xx}/\bar x^2_H$, $\langle\delta_x\delta\rangle=\xi_{x\delta}/\bar x_H$, and $\langle\delta\delta\rangle=\xi_{\delta\delta}$. 

The auto-correlation function of the isotropic $\HI$ spatial fluctuation field is~\cite{2004ApJ...608..622Z,2004HI}
\begin{equation}
\xi_{\psi\psi}=\xi_{xx}(1+\xi_{\delta\delta})+\bar x_H^2\xi_{\delta\delta}+\xi_{x\delta}(2\bar x_H+\xi_{x\delta})+\xi_{x\delta x\delta}.
\end{equation}
We should note that this $\HI$ auto-correlation function is only an approximation in that the three-point correlation terms neglected are generally substantial, as~\cite{2007ApJ...659..865L} pointed out. However, we mainly focused on the cross-correlation calculations, for which we did consider all of the higher order terms. Although the assumption that the density field is Gaussian is a reasonable approximation on most of the scales of interest, we create the ionizing field from a Poisson process, as we will discuss later, to account for its non-Gaussianity. We only use power spectrum to do the statistics so that the non-Gaussianity of the field is not captured~\cite{2004ApJ...608..622Z}. We neglect the redshift distortions and make use of the fact that the spin temperature is significantly higher than CMB at $z<$ 10~\cite{2011MNRAS.410.1377T}. We Fourier transform the correlation function, assuming that the quadratic terms are negligibly small. The $\HI$ power spectrum is
\begin{eqnarray}
P_{\psi\psi}^{(\rm{2h})}&=&P_{xx}^{(\rm{2h})}+\bar x_H^2P_{\delta\delta}^{(\rm{2h})}+2\bar x_HP_{x\delta}^{(\rm{2h})}+P_{x\delta x\delta}^{(\rm{2h})}\nonumber\\
&\simeq&P_{xx}^{(\rm{2h})}+\bar x_H^2P_{\delta\delta}^{(\rm{2h})}+2\bar x_HP_{x\delta}^{(\rm{2h})}.
\end{eqnarray}

The power spectrum can be calculated from a halo model by describing HII regions as bubbles. 
The two-halo term of $P_{x\delta x\delta}$ is higher order and negligible~\cite{2004HI}.

\begin{figure}
\includegraphics[width=10cm, height=9cm]{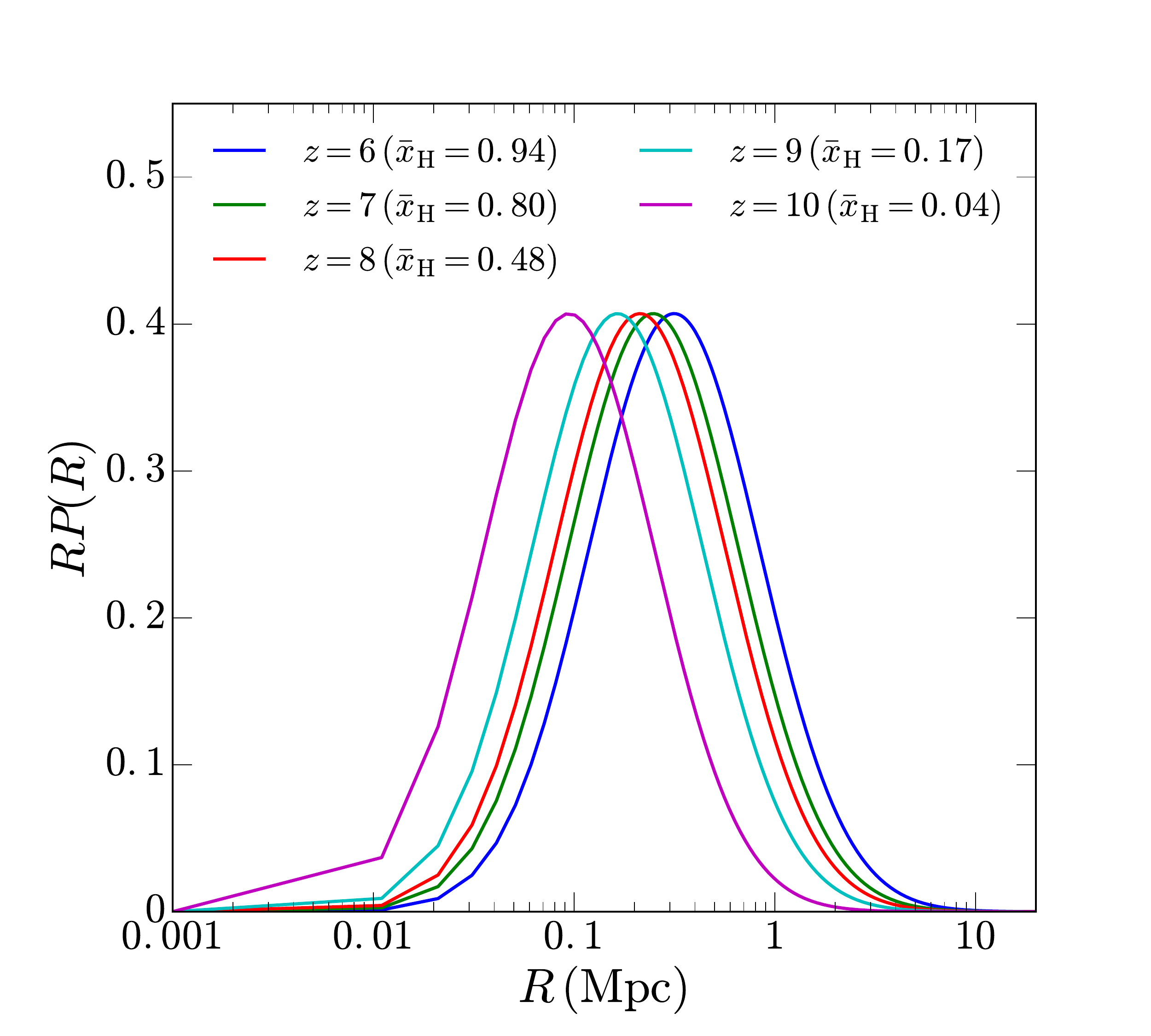}
\caption{Bubble size radius $R$ is assumed to satisfy a logarithmic distribution in Eq. (\ref{logp}). We show the distributions at redshifts $z$ = 6, 7, 8, 9, and 10. The mean ionization fractions $\bar x_{\rm{H}}$ at these redshifts are given in parentheses.}\label{pdfbub}
\end{figure}

The viral temperature of halo $T_{\rm{vir}}=5\times 10^4\,\rm{K}$ (suggested by~\cite{cora2}) sets the minimum halo mass
\begin{equation}
\frac{T_{\rm{vir}}}{10^4\rm{K}}=1.1\Big(\frac{\Omega_mh^2}{0.15}\Big)^{1/3}\Big(\frac{1+z}{10}\Big)\Big(\frac{M_{\rm{th}}}{10^8M_{\odot}}\Big)^{2/3}.
\end{equation}
With this threshold mass, we can calculate the mean number density of bubble $\bar n_b=\int_{M_{\rm{th}}}^{\infty}dM\frac{dn}{dM}$ and the average bubble size from $
\bar n_b=-(\ln\bar x_H)/\bar V_b$. When it is compared to the predicted value
\begin{equation}
\bar V_b=\int dR P(R)V_b(R)=\frac{4\pi \bar R^3}{3}e^{9\sigma^2_{\ln R}/2},\end{equation}
the bubble radius is constrained. The bubble radius $R$ is assumed to satisfy a logarithmic distribution~\cite{lidz21cm} as
\begin{equation}
P(R)=\frac{1}{R}\frac{1}{\sqrt{2\pi \sigma^2_{\ln R}}}e^{-\frac{1}{2}\Big(\frac{\ln(R/\bar R)}{\sigma_{\ln R }}\Big)^2}.\label{logp}
\end{equation}
In Figure~\ref{pdfbub}, we show the distribution of the bubble size at different redshifts.

Given the average volume and number density, the ionizing field is generated from a Poisson process
\begin{equation}
\langle x_e(\textbf{x})\rangle=1-e^{-n_b(\textbf{x})\bar V_b}, 
\end{equation}
and its number density is
\begin{equation}
n_b(\textbf{x})=\bar n_b (1+b\delta_L(\textbf{x})),
\end{equation}
where the density contrast $\delta_L$ is the matter density $\delta$ smoothed by a top-hat window of radius $R$. The top-hat window in Fourier space is
\begin{equation}
\bar W^n_R(k)=\frac{1}{\bar V^n_b}\int_0^{\infty}dR P(R)[V_b(R)W(kR)]^n.
\end{equation}
The shape factor for the ionizing field $x$ is defined as
\begin{equation}
X^{(x)}_l(k,M,z)=\bar x_H\ln{\bar x_H}b_{\rm{bubble}}\bar W_R(k)u_1\label{21shape},
\end{equation} with the bubble bias given by
\begin{equation}
b_{\rm{bubble}}=\frac{1}{\bar n_b}\int^{\infty}_{M_{\rm{th}}}b(M,z)\frac{dn}{d\ln M}\frac{dM}{M}.
\end{equation}
Here, $u_1$ is the Fourier transform of the NFW profile~\cite{nfw}; i.e., $u_1=M/\bar \rho_mu$. The NFW Fourier transform is
\begin{equation}
u(k,M,z)=\frac{1}{M}\int_0^{r_{\rm{vir}}}dr4\pi r^2\frac{\sin{kr}}{kr}\rho_{\rm{NFW}},
\end{equation}
which is derived from a standard NFW profile
\begin{equation}
\rho_{\rm{NFW}}=\rho_s\Big(\frac{r}{r_s}\Big)^{-1}\Big(1+\frac{r}{r_s}\Big)^{-2}.
\end{equation}
The detailed discussions of $\rho_s$, $r_s$ and $r_{\rm{vir}}$ can be found in Ref.~\cite{gamma16}. 

The 1-halo term of the $\HI$ field~\cite{Mortonson06,xiaomin05} is
\begin{eqnarray}
P_{\psi\psi}^{(\rm{1h})}&=&P_{xx}^{(\rm{1h})}+\bar x_H^2P_{\delta\delta}^{(\rm{1h})}+2\bar x_HP_{x\delta}^{(\rm{1h})}+P_{x\delta x\delta}^{(\rm{1h})}\nonumber\\
&\simeq&P_{xx}^{(\rm{1h})}+\bar x_H^2P_{\delta\delta}^{(\rm{1h})}+P_{x\delta x\delta}^{(\rm{1h})},
\end{eqnarray}
where $P_{xx}^{(\rm{1h})}$ = $(\bar x_e-\bar x_e^2)V_b\bar W_R^2$ and $P_{x\delta x\delta}^{(\rm{1h})}$ = $(\bar x_e-\bar x_e^2)\tilde P_{\delta\delta}$. The term $P_{x\delta}^{(\rm{1h})}$ is zero as we assume the bubble is completely ionized.
Here $\sigma_R^2$ = $\int dkk^2/2\pi^2\bar W_R^2(k)P_{\delta\delta}(k)$ and $\tilde P_{\delta\delta}$ = $P_{\delta\delta}V_b\sigma_R^2/\sqrt{P_{\delta\delta}^2+(V_b\sigma_R^2)^2}$. The two-halo term can be easily calculated with the shape factor in Eq. (\ref{21shape}).

On the other hand, the intensity mapping (IM) of $\ly$ emitters (LAEs) is a biased tracer of the same dark matter distribution, i.e., $\sim(1+\delta_{\eta})$. For simplicity, we use $\delta_{\ly}=\delta_{\eta}$. The cross-correlation between $\HI$ and LAEs is 
\begin{equation}
\xi_{\psi \eta}=\bar x_H(\xi_{\delta_{\eta}\delta}+\xi_{\delta_{\eta}x\delta})+\xi_{x\delta_{\eta}}, 
\end{equation}
and the 3D power spectrum is
\begin{equation}
P_{\psi \eta}=\bar x_H(P_{\delta_{\eta}\delta}+P_{\delta_{\eta}x\delta})+P_{x\delta_{\eta}}.
\end{equation}

Here, the two-halo and one-halo terms are given by
\begin{eqnarray}
P_{\psi \eta}^{(\rm{2h})}&=&\bar x_H(P_{\delta_{\eta}\delta}^{(\rm{2h})}+P_{\delta_{\eta}x\delta}^{(\rm{2h})})+P_{x\delta_{\eta}}^{(\rm{2h})}\nonumber\\
&\simeq&\bar x_H P_{\delta_{\eta}\delta}^{(\rm{2h})}+P_{x\delta_{\eta}}^{(\rm{2h})}
\end{eqnarray}
and
\begin{eqnarray}
P_{\psi \eta}^{(\rm{1h})}&=&\bar x_H(P_{\delta_{\eta}\delta}^{(\rm{1h})}+P_{\delta_{\eta}x\delta}^{(\rm{1h})})+P_{x\delta_{\eta}}^{(\rm{1h})}\nonumber\\
&\simeq&0,
\end{eqnarray}
respectively. The subscript $\delta_{\eta}x\delta$ essentially refers to $\delta_{\eta},x\delta$ and the ``," is omitted for simplicity.

On small scales $P^{1h}_{\delta_{\eta}x\delta}$ cancels out the term $P^{1h}_{\delta_{\eta}\delta}$, so the summation is almost zero~\cite{lidz21cm}. Also, the large-scale information of $P^{2h}_{\delta_{\eta}x\delta}$ should be very negligible. With all of these approximations, the final power spectrum of the $\HI$\mbox{--}$\ly$ cross-correlation is
\begin{equation}
P_{\psi\eta}\simeq P_{x\delta_{\eta}}^{(\rm{2h})}+\bar x_HP^{(\rm{2h})}_{\delta_{\eta}\delta}.\label{cross}
\end{equation}
The halo-model approach, i.e.,
\begin{equation}
P^{1h,XY}(k,z)=\int dM\frac{dn}{dM}X_l(k,M,z)Y_l(k,M,z)\label{1h}
\end{equation}
and
\begin{eqnarray}
P^{2h,XY}(k,z)&=&P_{\rm{lin}}(k,z)\int dM\frac{dn}{dM}b(M,z)\tilde X_l(k,M,z)\nonumber\\
&&\times\int dM\frac{dn}{dM}b(M,z)\tilde Y_l(k,M,z),\label{2h}
\end{eqnarray}
can be used to calculate each power spectrum in Eq. (\ref{cross}). In these equations, $dn/dM$ is the mass function and $b(M,z)$ is the bias. The linear matter power spectrum 
is $P_{\rm{lin}}$. We will work out the shape factors $X_l(k,M,z)$ and $Y_l(k,M,z)$ (or $\tilde X_l(k,M,z)$ and $\tilde Y_l(k,M,z)$) for $\ly$ in the next section.

\section{$\ly$ emission}
\label{lya}

The UV radiation emitted from massive and short-lived stars can ionize the neutral hydrogen in the interstellar medium (ISM) in galaxies and the number of ionizing photons closely depends on the star formation rate (SFR). In this work we consider an SFR model that is consistent with numerical simulations. The fitted SFR~\cite{silva_lya} is
\begin{eqnarray}
\frac{{\rm SFR}(M,z)}{M_{\odot} \rm{yr}^{-1}}&=&2.8\times10^{-28}M^a\Big(1+\frac{M}{M_1}\Big)^{b}\nonumber\\
&&\times\Big(1+\frac{M}{M_2}\Big)^{d},\nonumber\\
\end{eqnarray}
where $a=2.8$, $b=-0.94$, $d=-1.7$, $M_1=10^9M_{\odot}$, and $M_2=7\times 10^{10}M_{\odot}$.

The ionizing photons could escape the galaxies with a fraction $f_{\rm{esc}}(M,z)=e^{-\alpha(M/M_{\odot})^{\beta}}$, but the remains will ionize the hydrogen and 66\% of the ionization will result in a recombination process that produces $\ly$ photons. The dust in the ISM can also absorb the $\ly$ emissions, and the remaining fraction that survives the dust extinction is $f_{\rm{Lya}}(z)$. The luminosity due to the recombination is then calculated as
\begin{equation}
L_{\rm{rec}}^{\rm{GAL}}(M,z)=1.55\times 10^{42}(1-f_{\rm{esc}})f_{\rm{Lya}}\frac{\rm{SFR}}{M_{\odot} \rm{yr}^{-1}}\,(\rm{erg}\,\rm{s}^{-1}).
\end{equation}

The ionizing radiation can heat the gas so that the process of hydrogen excitation and cooling produces $\ly$ emission as well. The luminosities due to excitation and cooling are
\begin{equation}
L_{\rm{exc}}^{\rm{GAL}}(M,z)=4.03\times 10^{41}(1-f_{\rm{esc}})f_{\rm{Lya}}\frac{\rm{SFR}}{M_{\odot} \rm{yr}^{-1}}\,(\rm{erg}\,\rm{s}^{-1}),
\end{equation}
and 
\begin{eqnarray}
L_{\rm{cooling}}^{\rm{GAL}}(M,z)&=&1.69\times 10^{35}f_{\rm{Lya}}(1+\frac{M}{10^8})\nonumber\\
&&\times(1+\frac{M}{2\times 10^{10}})^{2.1}(1+\frac{M}{3\times 10^{11}})^{-3}\,(\rm{erg}\,\rm{s}^{-1}),\nonumber\\
\end{eqnarray}
respectively.

Besides these line emissions, the continuum produces $\ly$ photons through stellar radiation, free-free (ff), free-bound (fb), and two-photon (2$\gamma$) processes. Among these contributions, the stellar emission with a blackbody spectrum below the Lyman limit is dominant and its luminosity is
\begin{equation}
L_{\rm{stellar}}^{\rm{GAL}}(M,z)=5.12\times 10^{40}f_{\rm{Lya}}\frac{\rm{SFR}}{M_{\odot} \rm{yr}^{-1}}\,(\rm{erg}\,\rm{s}^{-1}).
\end{equation}
Our calculation takes all of these continuum lines into account, and the detailed line luminosity can be found in Ref.~\cite{silva_lya}.

The total $\ly$ luminosity $L(M,z)$ from a galaxy is a summation of all of the above components and the shape factor for $\ly$ field is
\begin{equation}
X_l(k,M,z)=\frac{L(M,z)}{4\pi D_L^2}yD^2_Au(k,M,z).
\end{equation}
Here the conversion factor from frequency to comoving distance is $y$ = $d\chi/d\nu$ = $\lambda/(ca^2)d\chi/dz$, $\lambda$ is the line rest frame wavelength, and $D_L$ and $D_A$ are luminosity and angular comoving distances, respectively. 

This construction of shape factor is only a mathematical definition that facilitates the halo-model calculations in Eqs. (\ref{1h}) and (\ref{2h}). We should note that for individual point-like $\ly$ emitters, it would appear to be extended due to spatial diffusion of $\ly$ photons~\cite{2010ApJ...716..574Z,2011ApJ...726...38Z}. Also, the scattering of photons on the red-side of $\ly$ in the IGM would further damp the $\ly$ flux along the line of sight~\cite{1998ApJ...501...15M}. We first consider that the $\ly$ emission is a biased tracer of the underlying dark matter distribution and phenomenologically account for the extended structure by the mass- and redshift-dependent quantity $b(M,z)$ in Eq. (\ref{2h}) and the luminosity function $L(M,z)$. Next we will discuss some dominating effects of IGM on the $\ly$ emissions, but effects such as the damping wing of $\ly$ have to rely on a numerical simulation.

The mean $\ly$ intensity varies at different redshifts as
\begin{equation}
\bar I_{\rm{Lya}}(z)=\int_{M_{\rm{min}}}^{M_{\rm{max}}}dM\frac{dn}{dM}\frac{L(M,z)}{4\pi D_L^2}yD^2_A,
\end{equation}
where $M_{\rm{min}}=10^8 M_{\odot}$ and $M_{\rm{max}}=10^{13} M_{\odot}$.

\begin{figure}
\includegraphics[width=8cm, height=7cm]{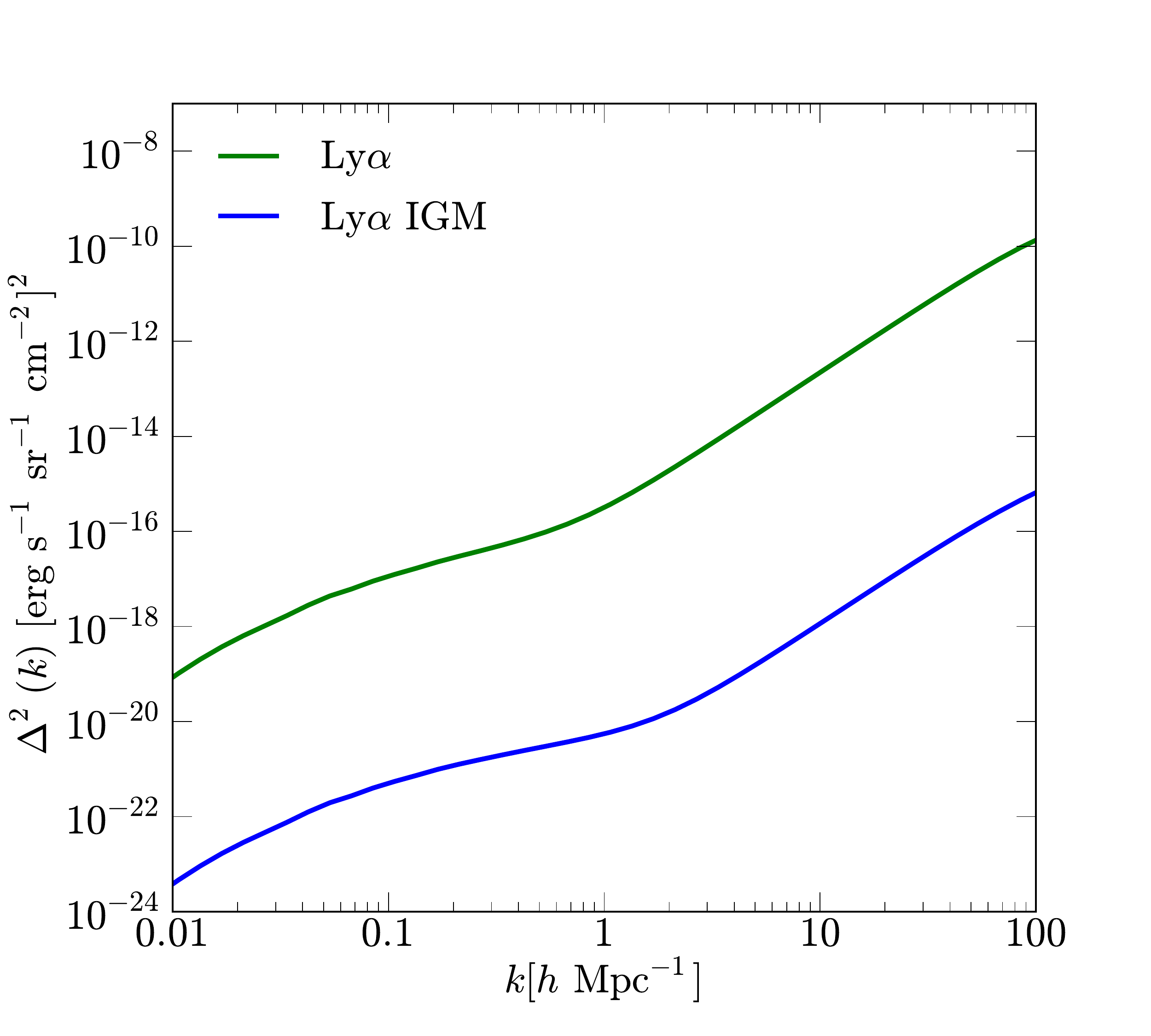}
\includegraphics[width=8cm, height=7cm]{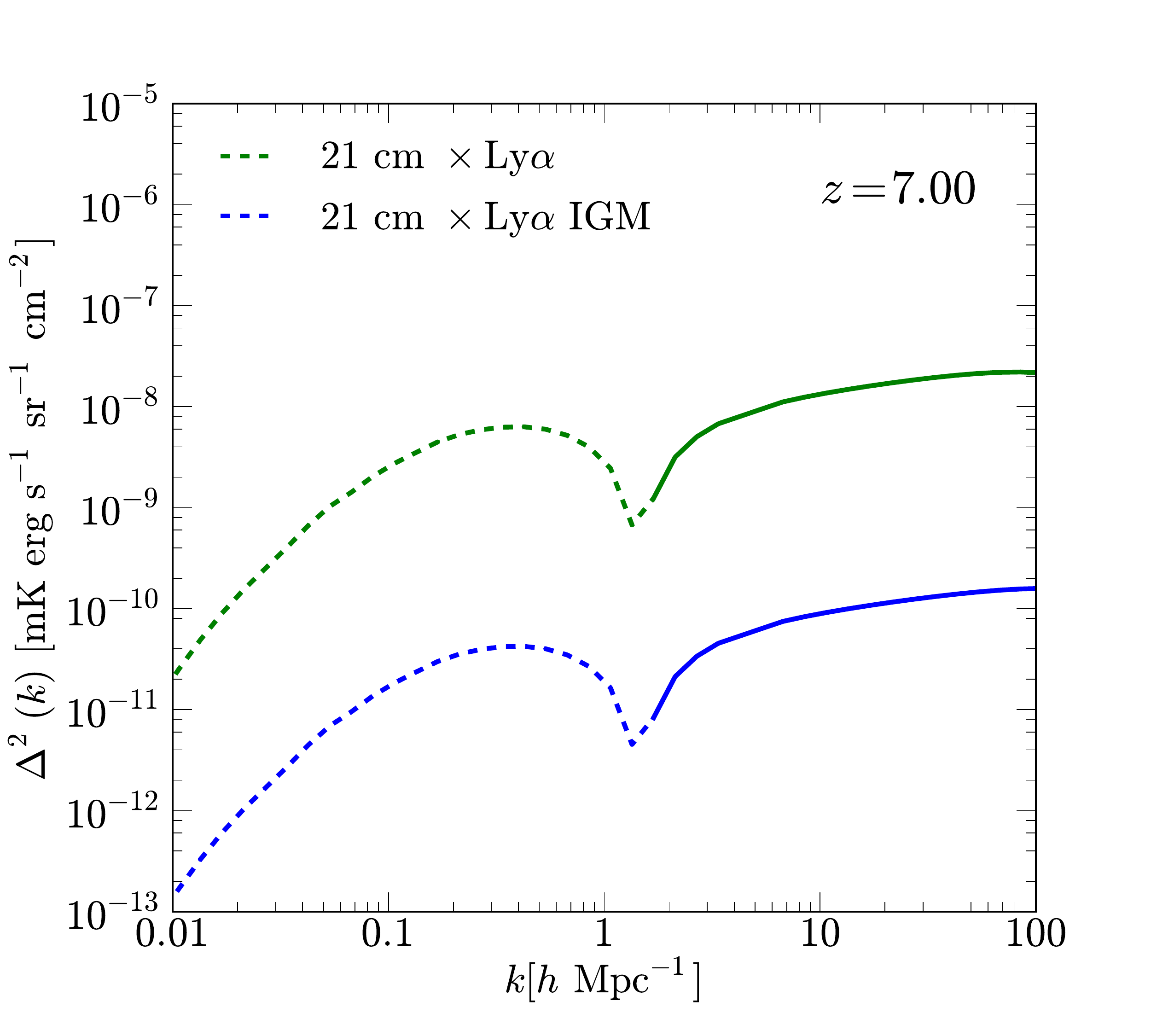}
\caption{IGM contribution to both the auto- and cross-power spectra at $z$ = 7. The IGM component is negligible, compared to the galaxy. The dashed portion is negative. In the $y$ axis, $\Delta^2(k)$ = $k^3/(2\pi^2)P(k)$.}\label{galvsIGM}
\end{figure}

The escaped photons from galaxies can ionize the IGM, which can also emit $\ly$ photons due to the recombination process. The recombination rate is
\begin{equation}
\dot n_{\rm{rec}}=\alpha n_en_{\rm{HII}},
\end{equation}
where $n_e=x_en_b$, $n_{\rm{HII}}=x_en_bC$, $C=(1-Y_{\rm p})/(1-3Y_{\rm p}/4)$, and $\alpha$ is a case A comoving recombination coefficient. The luminosity function of IGM is $L^{\rm{IGM}}_{\rm{rec}}=f_{\rm{rec}}\dot n_{\rm{rec}}E_{\rm{Ly}\alpha}$ and the fraction $f_{\rm{rec}}$ is spin-temperature dependent. We show the contribution of IGM in Figure \ref{galvsIGM} and it is seen that the IGM contribution is negligible for both auto- and cross-power spectra.

Another IGM contribution to $\ly$ emission comes from the scattering of Ly$n$ photons escaping from galaxies. From the previous calculations~\cite{silva_lya,silva2016,pullen13}, it is found that the diffuse IGM contribution is a few orders of magnitude smaller than the galaxies. Therefore, we ignore this contribution to the overall $\ly$ signal.

\section{Low-z Foregrounds}
\label{lowzfg}

The $\ly$ emission at the EoR can be significantly contaminated by low-$z$ foregrounds. The foreground at $z_f$ projected onto the source plane $z_s$ becomes anisotropic as the wave vector of the foreground power spectrum in Fourier space becomes $\textbf{k}_{f\rightarrow s}=(\chi_s/\chi_f k_{\perp},\,y_s/y_fk_{\parallel})$, which is not radially symmetric. The low-$z$ foregrounds are identified as $\Ha$ [$6563\,\AA$, $z$ = 0.5], OIII [$5007\,\AA$, $z$ = 0.9], and OII [$3727\,\AA$, $z$ = 1.6] with luminosities $L_{H\alpha}=1.3\times 10^{41}\frac{\rm{SFR}}{M_{\odot} \rm{yr}^{-1}}$, $L_{OII}=7.1\times 10^{40}\frac{\rm{SFR}}{M_{\odot} \rm{yr}^{-1}}$, and $L_{\rm{OIII}}=1.3\times 10^{41}\frac{\rm{SFR}}{M_{\odot} \rm{yr}^{-1}}$, respectively. The low-$z$ SFR is exclusively modeled as
\begin{equation}
\frac{{\rm SFR}(M, z)}{M_{\odot} \rm{yr}^{-1}}=10^{a+bz}\Big(\frac{M}{M_1}\Big)^c\Big(\frac{M}{M_2}\Big)^d
\end{equation}
for the foreground line emissions. This SFR model is fitted to the numerical simulations below $z$ = 2 and the parameters are constrained as $a$ = $-9.097$, $b$ = 0.484, $c$ = 2.7, $d$ = $-4.0$, $M_1$ = $10^8M_{\odot}$, and $M_2$ = $8\times 10^{11}M_{\odot}$~\cite{yanlya}.

\begin{figure}
\includegraphics[width=7.5cm, height=6cm]{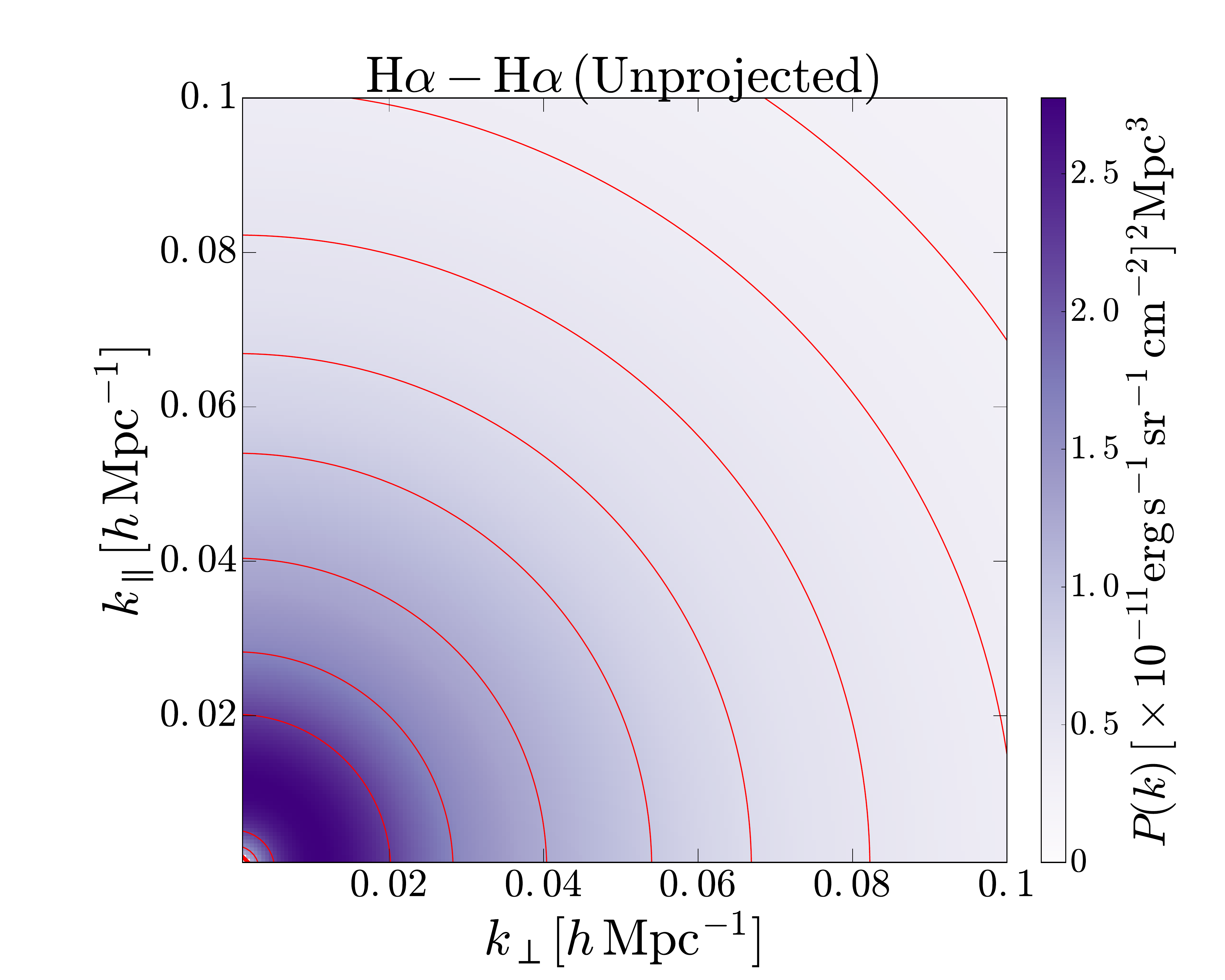}
\includegraphics[width=7.5cm, height=6cm]{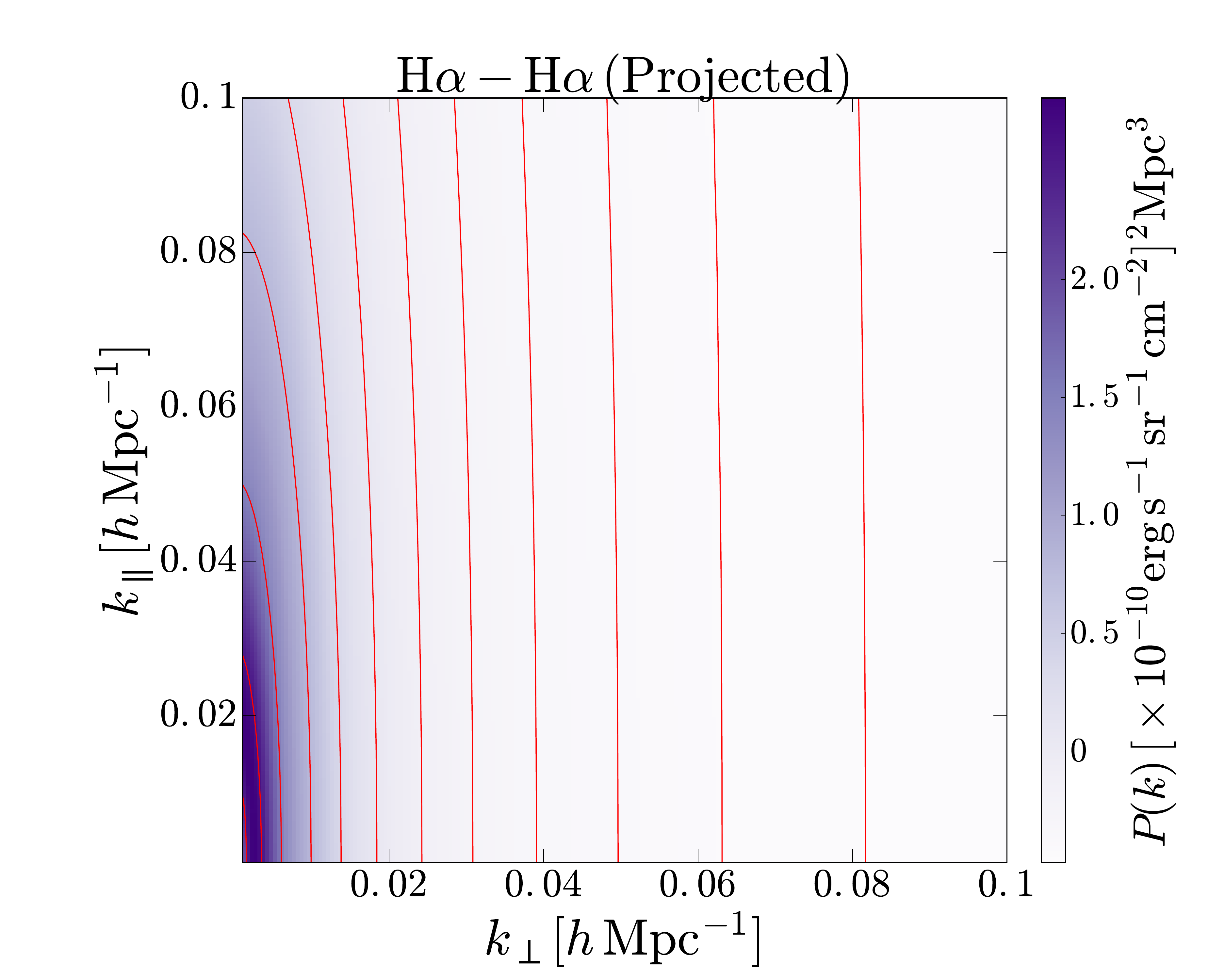}
\caption{Power spectrum projection of $\Ha$ at $z$ = 7. OIII and OII power spectrum projections have similar patterns.}\label{proj}
\end{figure}

\begin{figure}
\includegraphics[width=8cm, height=7cm]{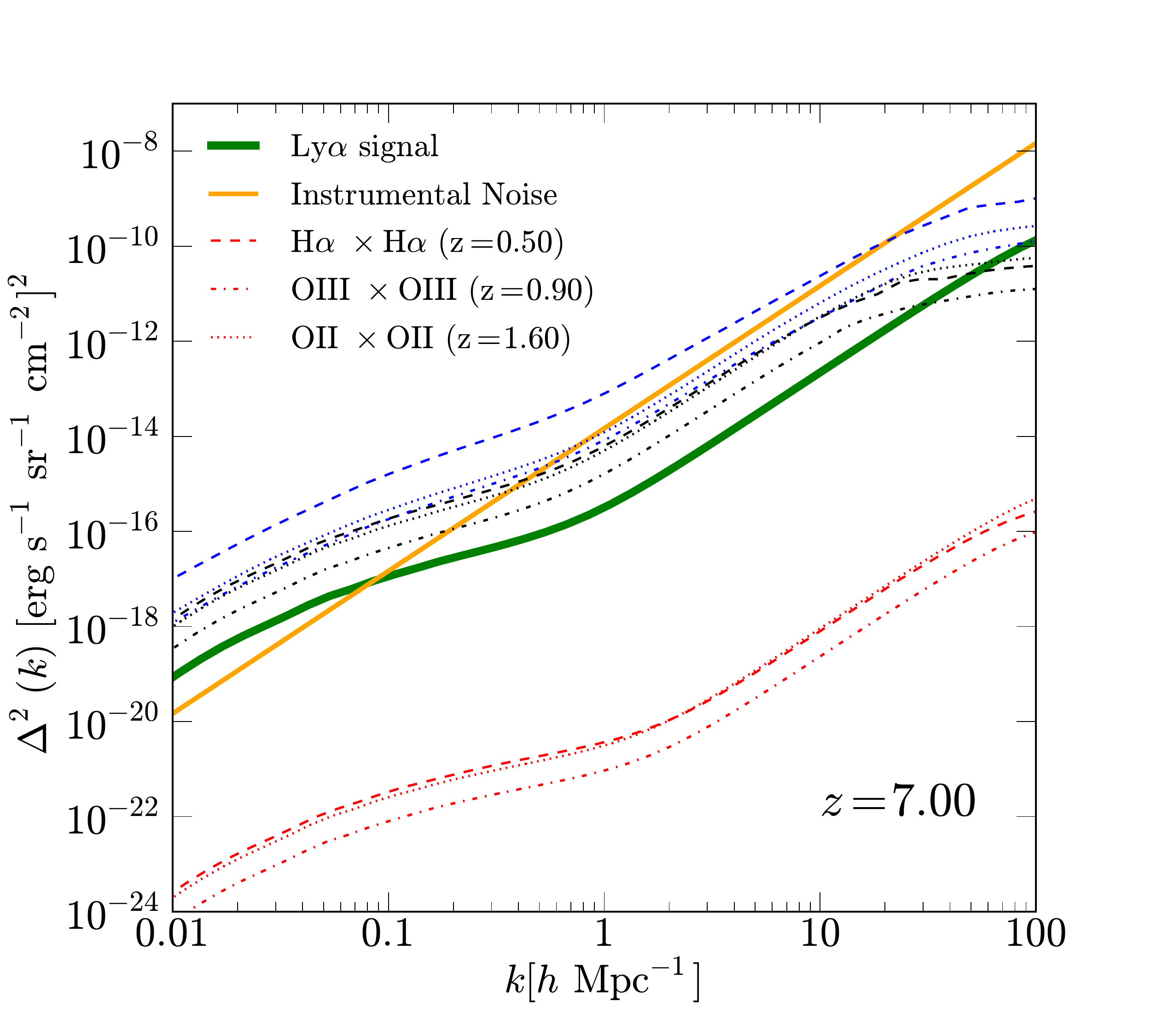}
\caption{$\ly$ power spectrum at $z$ = 7 (green). We show the low-$z$ foreground power spectra with no projection (black), projection (blue), and masking (red). The instrumental noise (orange) is derived from the proposed CDIM specification listed in Table~\ref{t2}.}\label{slyaat7}
\end{figure}

The projected power spectrum of the foreground is then expressed as
\begin{equation}
P_{f\rightarrow s}(k_{\perp},k_{\parallel},z_s)=\Big(\frac{\chi_s}{\chi_f}\Big)^2\frac{y_s}{y_f}P(k_f,z_f)
\end{equation}
and $k_f$ = $\sqrt{(\chi_s/\chi_f)^2k^2_{\perp}+(y_s/y_f)^2k^2_{\parallel}}$. In Figure \ref{proj}, we show the unprojected and projected power spectra for $\Ha$. OIII and OII show similar contours so  they are neglected. In Figure \ref{slyaat7}, the blue and red curves are radially averaged from the anisotropic power spectra. 

\begin{figure}
\includegraphics[width=8cm, height=7cm]{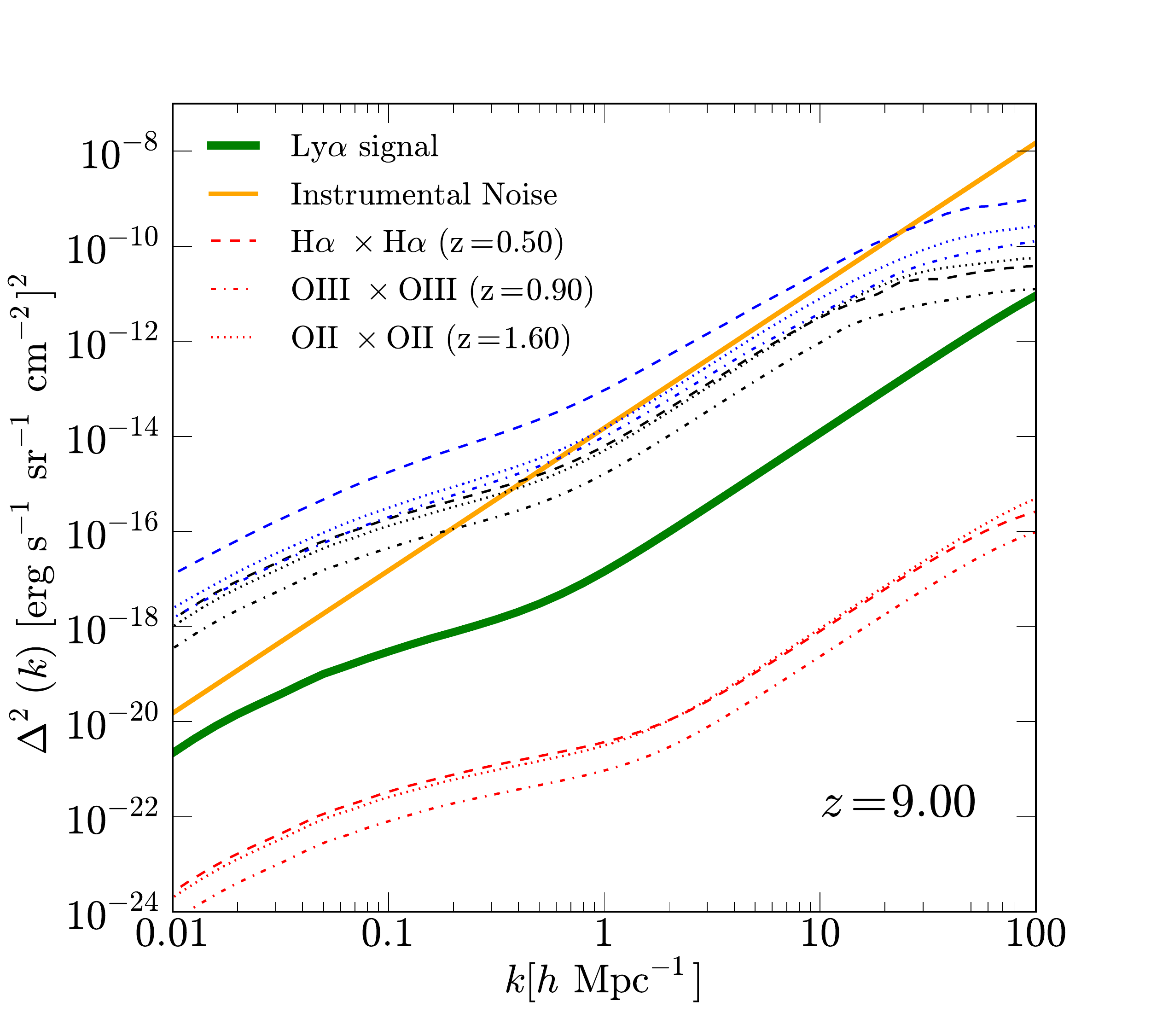}
\caption{$\ly$ power spectrum at $z$ = 9 (green). The description of other lines is the same as Figure \ref{slyaat7}.}\label{slyaat9}
\end{figure}

\begin{figure}
\includegraphics[width=8cm, height=7.2cm]{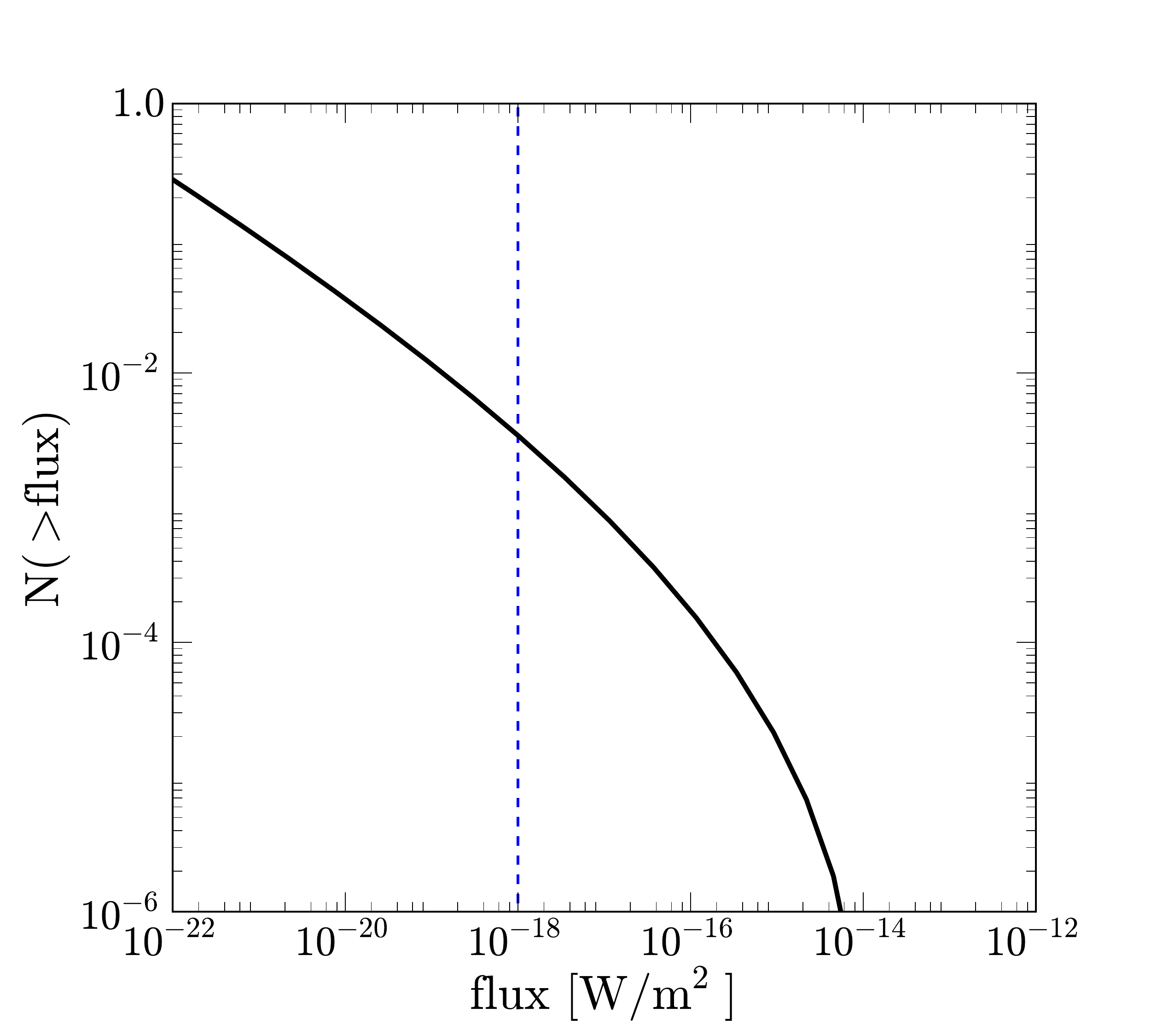}
\caption{Flux threshold of $\Ha$ line at $z\,=\,0.5$.}\label{fluxmask}
\end{figure}

In Figures \ref{slyaat7} and \ref{slyaat9}, we show the power spectra for $\ly$ and the foreground lines $\Ha$, OIII, and OII with projection and with flux masking at redshifts $z$ = 7 and 9. The projected foreground emissions are much higher than the $\ly$ lines. By selecting the brightest sources at the flux detection threshold and forming a mask, we can effectively remove those ``hot" pixels which only account for a very tiny fraction of the sky coverage~\cite{pullen13,yanlya}.  As can be seen in Figure~\ref{fluxmask}, we show the percentage of the removed pixels as the threshold flux changes. We find that a flux cut at $10^{-18}\rm{W/m^2}$ can significantly lower amplitudes of the foreground power spectra while only removing less than 0.1\% of the pixels. Therefore, the flux masking procedure makes the low-$z$ foregrounds negligible.

\begin{figure}
\includegraphics[width=8cm, height=7cm]{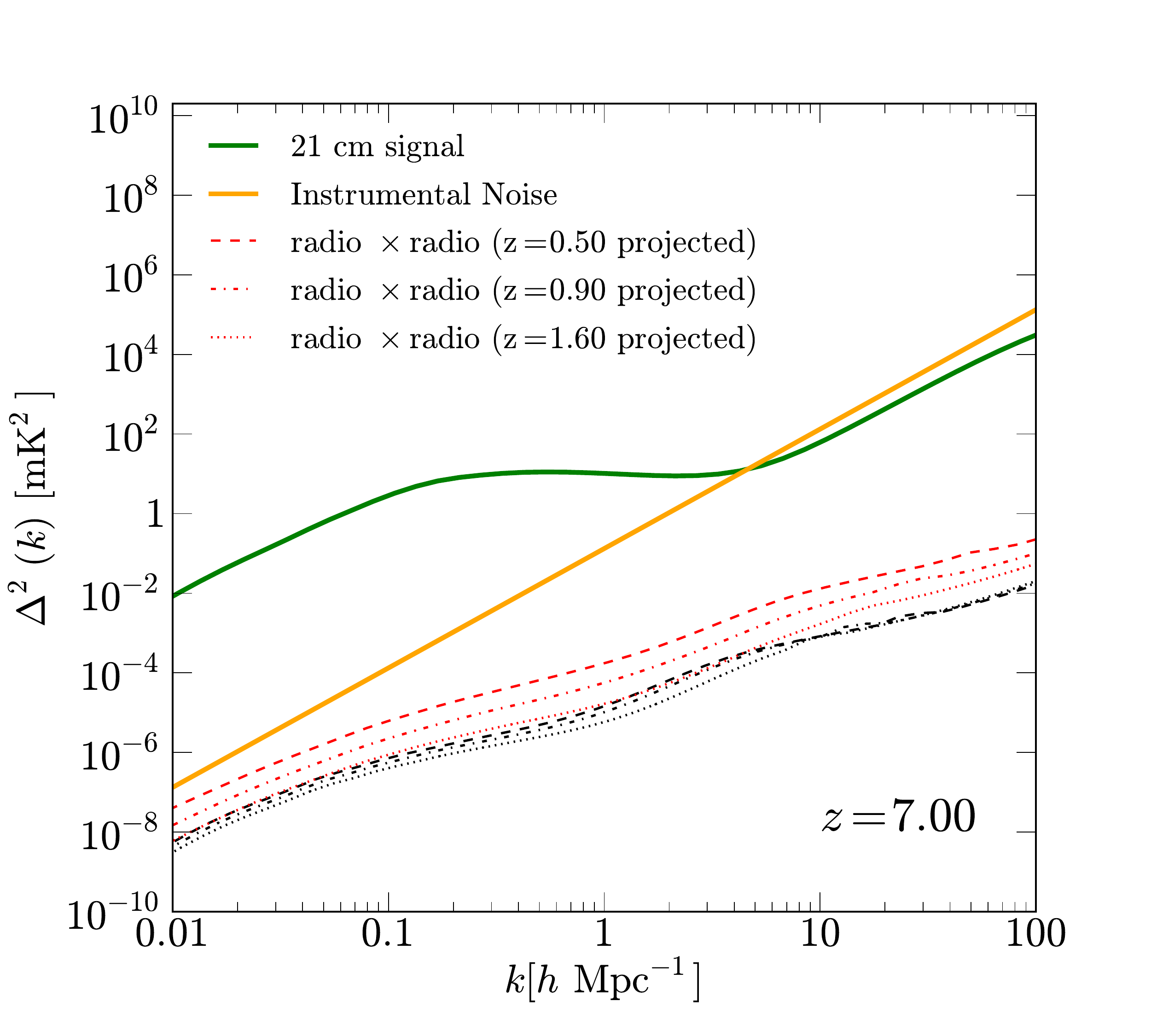}
\caption{$\HI$ power spectrum at $z$ = 7 (green). The radio flux cut is $S_{\rm{cut}}=1\,\rm{mJy}$. The instrumental noise (orange) is derived from the SKA specification in Table~\ref{t1}. The low-$z$ foreground line and its projection are shown in black and red, respectively, for each redshift when a $\ly$ foreground line is present. A spectral fitting scheme for the radio sources is assumed and the suppression factor is assumed to be $10^{-6}$~\cite{aliu_ps}.}\label{s21at7}
\end{figure}

\begin{figure}
\includegraphics[width=8cm, height=7cm]{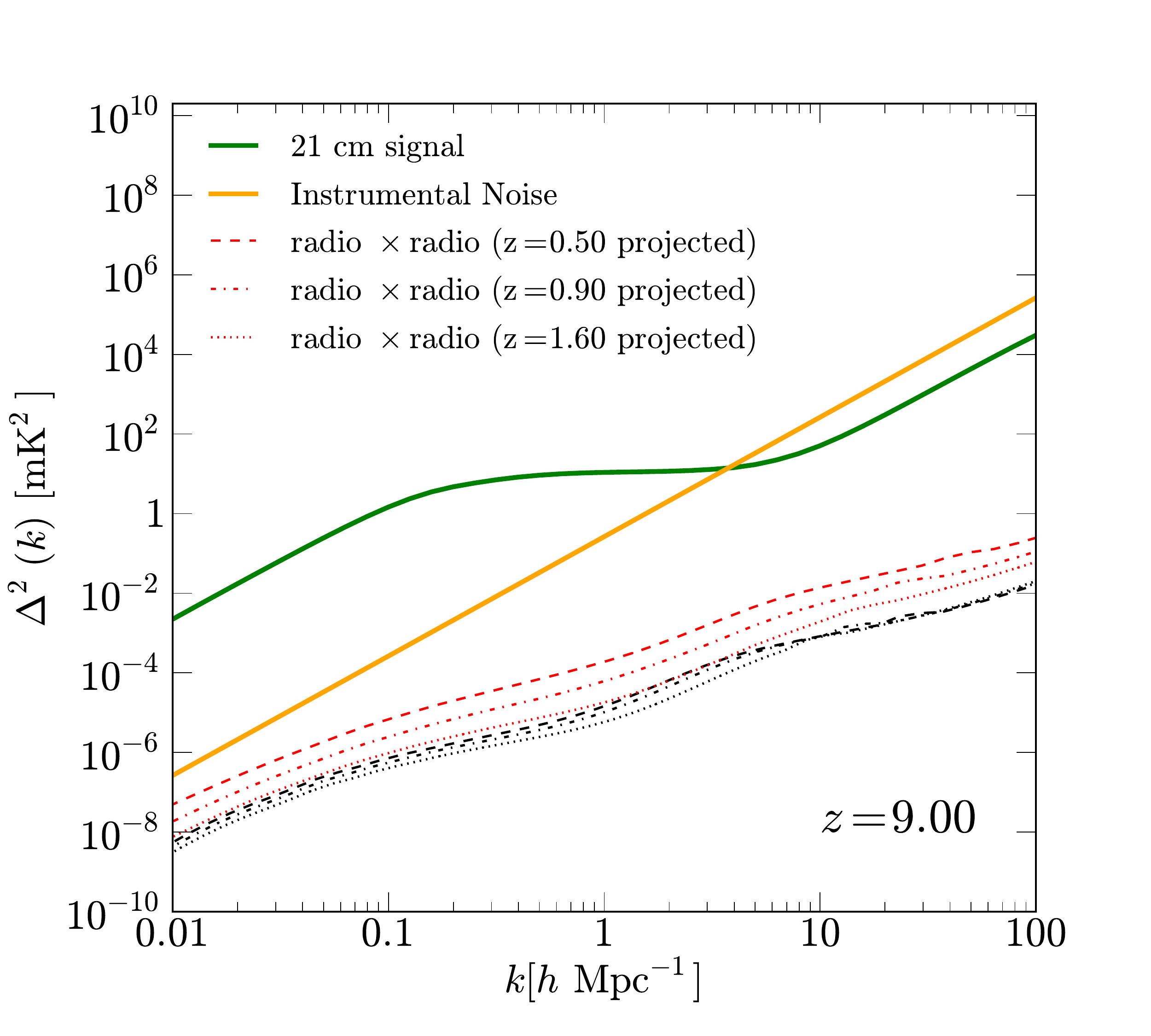}
\caption{$\HI$ power spectrum at $z$ = 9 (green). The description of other lines is the same as Figure \ref{s21at7}.}\label{s21at9}
\end{figure}

\begin{figure}
\includegraphics[width=8cm, height=7cm]{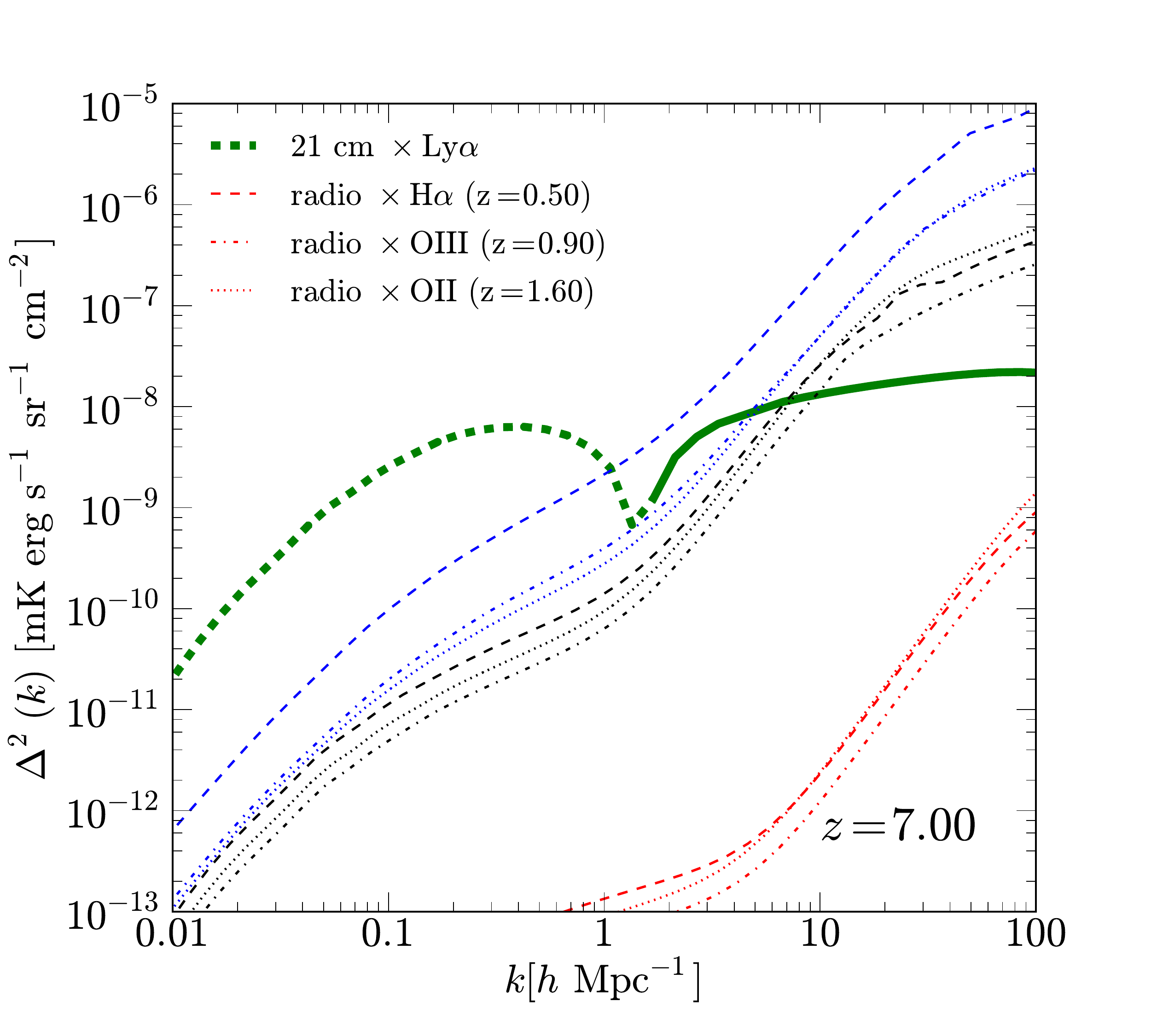}
\caption{$\HI$\mbox{--}$\ly$ cross-power spectrum at $z$ = 7 (green); the green dashed line indicates that it is negative. We show the radio-line foreground cross-power spectra with no projection (black), projection (blue), and masking (red). The radio flux cut is $S_{\rm{cut}}=1\,\rm{mJy}$. The instrumental noise for the cross-correlation is zero.}\label{scrossat7}
\end{figure}

\begin{figure}
\includegraphics[width=8cm, height=7cm]{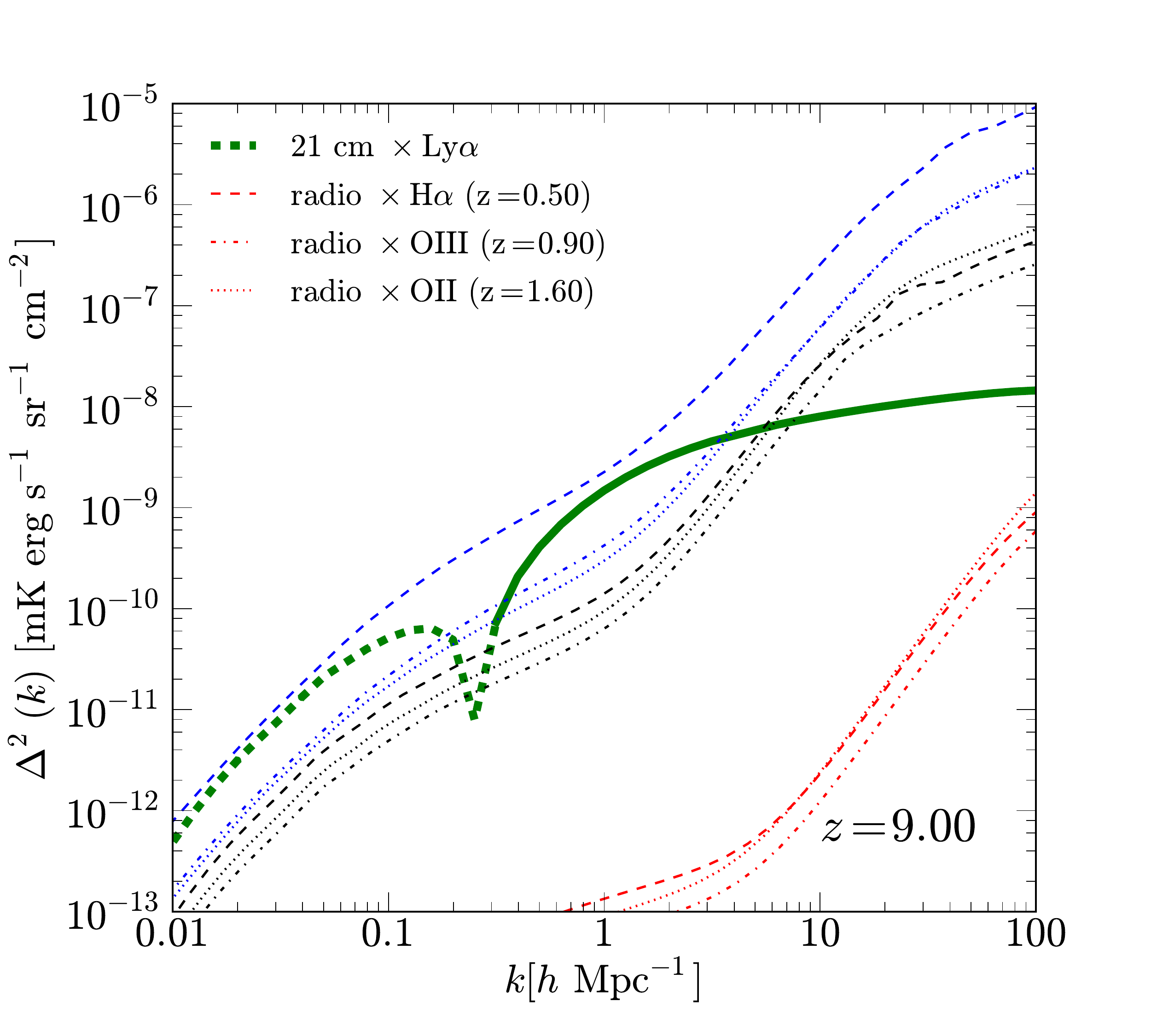}
\caption{$\HI$\mbox{--}$\ly$ cross-power spectrum at $z$ = 9 (green). The description of other lines is the same as Figure \ref{scrossat7}.}\label{scrossat9}
\end{figure}

\begin{figure}
\includegraphics[width=8cm, height=7cm]{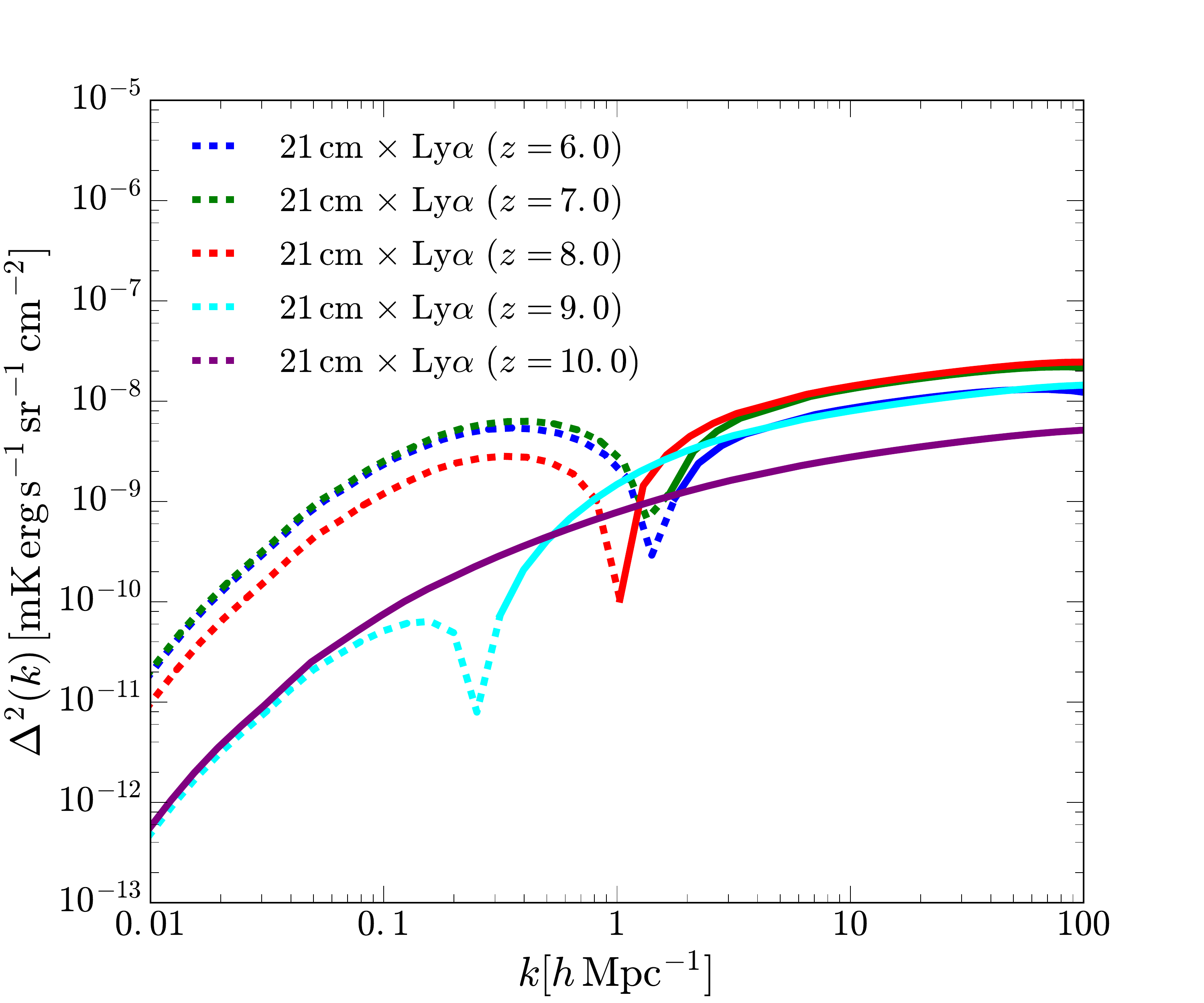}
\includegraphics[width=8cm, height=7cm]{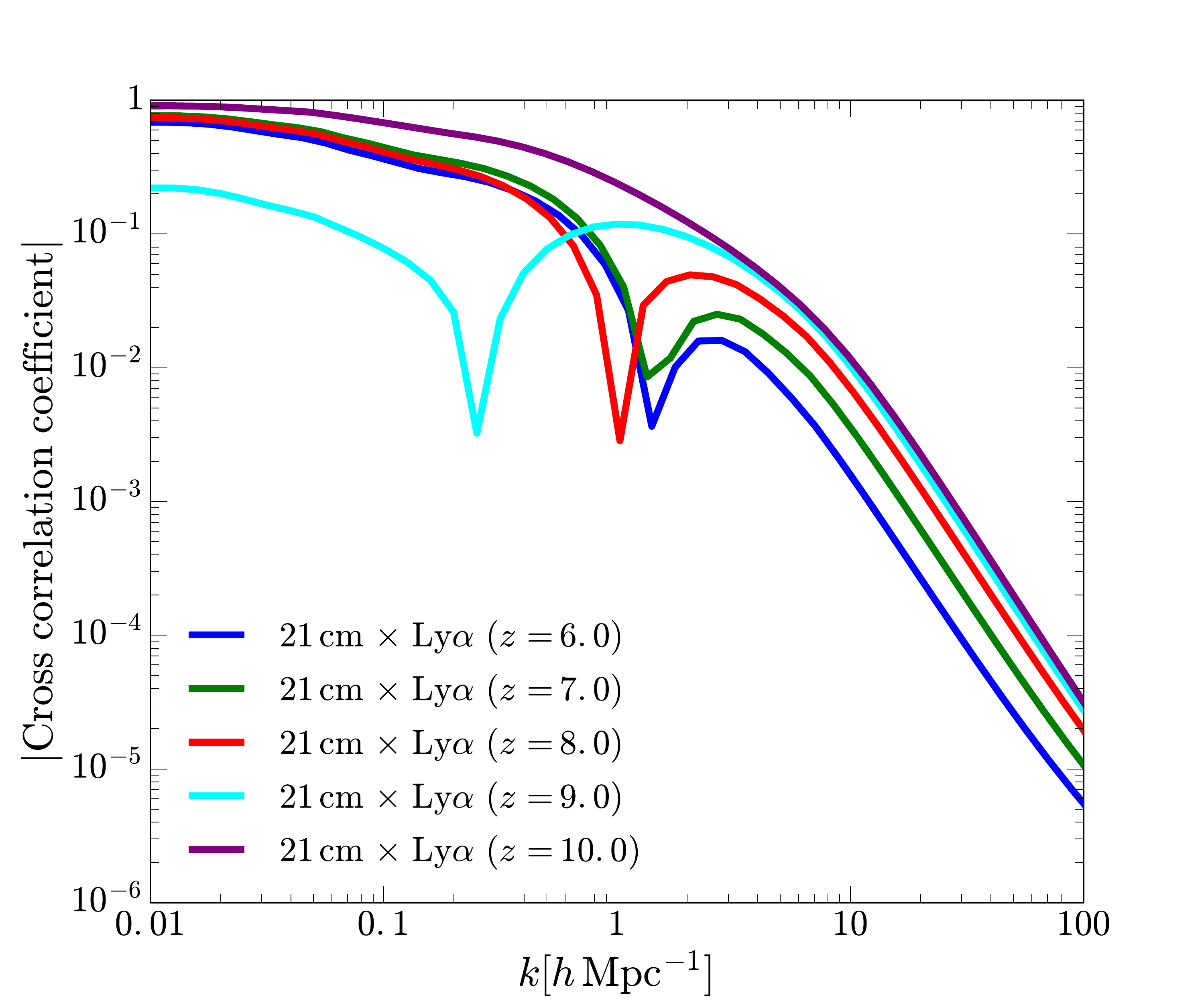}
\caption{Top: redshift evolution of the $\HI$\mbox{--}$\ly$ cross-correlation as a function of $k$. The mean ionization fractions are $\bar x_{\rm{H}}$ = 0.94 ($z$ = 6), $\bar x_{\rm{H}}$ = 0.80 ($z$ = 7), $\bar x_{\rm{H}}$ = 0.48 ($z$ = 8), $\bar x_{\rm{H}}$ = 0.16 ($z$ = 9), and $\bar x_{\rm{H}}$ = 0.04 ($z$ = 10). Bottom: cross-correlation coefficient as a function of $k$ at different redshifts.}\label{evol}
\end{figure}

The synchrotron radiation dominates the $\HI$ signals but its smooth spectral feature can easily be used to isolate this component in frequency domain, also the galactic foregrounds are not correlated with extragalactic line emissions at low-$z$. So we do not expect any noticeable cross-correlations between galactic synchrotron and $\ly$ foregrounds. However, the radio point sources that are too faint to be resolved are indeed correlated with the low-$z$ foregrounds within $0.5<z<1.6$, so this component would be picked up in the $\HI$\mbox{--}$\ly$ cross-correlation, making it not as systematic-free as expected. To estimate its contribution, we use the model in Refs.~\cite{radio21cm,radiomodel, paoloradio}. The model is described as
\begin{equation}
\delta_{\rm{radio}}(\textbf{x})=\Big(\frac{\partial B_{\nu}}{\partial T}\Big)^{-1}I_{\nu}^{\rm{radio}}\delta_{g}(\textbf{x})
\end{equation}
based on the fact that the radio point source is a tracer of underlying density field. Here we have defined $I_{\nu}^{\rm{radio}}$ = $\int_0^{S_{\rm{cut}}}dS SdN/dS$ and assume a flux limit $S_{\rm{cut}}$ = 1$\,\rm{mJy}$, above which the radio point sources are bright enough to be resolved. The flux distribution is a simple power-law, i.e., $dN/dS\, =\, A(S/S_0)^{\alpha}$, where $A$ = 4$\,\rm{mJy}^{-1}\,\rm{sr}^{-1}$, $S_0$ = 880$\,\rm{mJy}$, and $\alpha\,=\,-1.75$~\cite{aliu_ps}. Also, $\partial B_{\nu}/\partial T$ = $99.27\,\rm{Jy}\,\rm{sr}^{-1}/(\mu K)$$x^4e^x$$/(e^x$-$1)^2$ and $x$ = $h\nu/k_bT_{\rm{CMB}}$ = $\nu/56.84\,(\rm{GHz})$.
In Fourier space, the shape functions of the halo model for the radio sources are described as
\begin{equation}
X_l(k,M,z)=\frac{\sqrt{2N_cN_su(k,M,z)+N_s^2u^2(k,M,z)}}{\bar n_g}
\end{equation}
and
\begin{equation}
\tilde X_l(k,M,z)=\frac{N_su(k,M,z)}{\bar n_g},
\end{equation}
which are directly inserted into Eqs. (\ref{1h}) and (\ref{2h}) to obtain the one-halo and two-halo terms for the point-source clustering. Both $X_l(k,M,z)$ and $\tilde X_l(k,M,z)$ describe the Fourier-space profile of a point source with mass $M$ located at redshift $z$.
Here the central and satellite galaxy numbers are
\begin{equation}
N_c(M)=\frac{1}{2}\Big[1+{\rm erf}\Big(\frac{\log_{10}M-\log_{10}{M_{{\rm min}}}}{\sigma_M}\Big)\Big]
\end{equation}
and 
\begin{equation}
N_s(M)=\frac{1}{2}\Big[1+{\rm erf}\Big(\frac{\log_{10}M-\log_{10}{2M_{{\rm min}}}}{\sigma_M}\Big)\Big]\Big(\frac{M}{M_s}\Big)^{\alpha_s}.
\end{equation}
The mean galaxy number density is
\begin{equation}
\bar n_g(z)=\int dM n(M,z)N_g(M),
\end{equation}
where $N_g(M)=N_c(M)+N_s(M)$. The parameters determined from luminosity and color dependence of galaxy clustering in the SDSS DR7 main galaxy sample are $M_{\rm{min}}$ = $10^{9}M_{\odot}$, $\sigma_{M}$ = 0.2, $M_s$ = $5\times10^{10}M_{\odot}$, and $\alpha_s$ = 1~\cite{sdssHOD}.

We estimate that the radio foreground contributions at the $\ly$ foreground redshifts and the raw power spectra are a few orders of magnitude higher than the $\HI$ signal as revealed by~\cite{aliu_ps,2014MNRAS.444.3183A}. Therefore, the foreground suppression is very crucial; the spectral fitting procedure studied in~\cite{aliu_ps} demonstrated that the radio foregrounds can be reduced by six orders of magnitude in map space, and it has been validated that this is true from flux cut 0.1--100 \rm{mJy}. Consequently, the radio foreground contamination becomes negligible, and we show all of the power spectra in Figures \ref{s21at7} and \ref{s21at9} at redshifts $z$ = 7 and 9. We see that the resulting radio point sources have very negligible contaminating power on the $\HI$ measurements. Finally,  we show the $\HI$\mbox{--}$\ly$ cross-power spectra at redshifts $z$ = 7 and 9 in Figures \ref{scrossat7} and \ref{scrossat9} with both foreground separation schemes incorporated. In Figure~\ref{evol}, we show both the evolutions and cross-correlation coefficients of the $\HI$\mbox{--}$\ly$ cross-power spectrum as a function of $k$ from $z$ = 6 to $z$ = 10.

Despite the fact that all of the foreground cross-correlations are small, this has to rely on the assumption that we have very efficient foreground mitigation strategies for both $\HI$ and $\ly$ measurements. Non-negligible foreground residuals on $\HI$ and $\ly$ would make a great impact on the power spectrum uncertainties, even if they are uncorrelated.

\begin{figure}
\includegraphics[width=7.5cm, height=6cm]{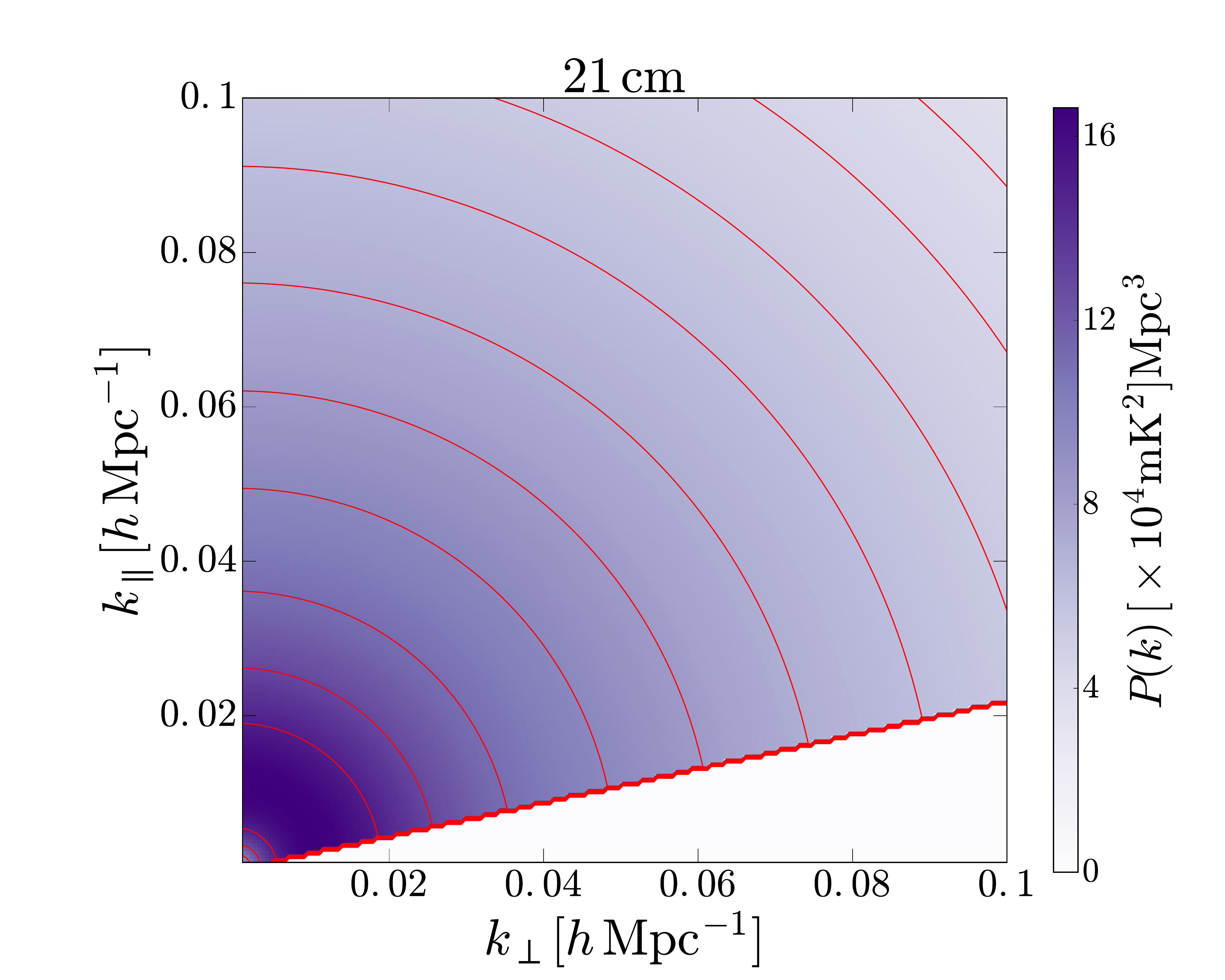}
\caption{2D power spectrum of $\HI$ at $z$ = 7 with foreground wedge.}\label{wedge}
\end{figure}

\begin{figure}
\includegraphics[width=8cm, height=7cm]{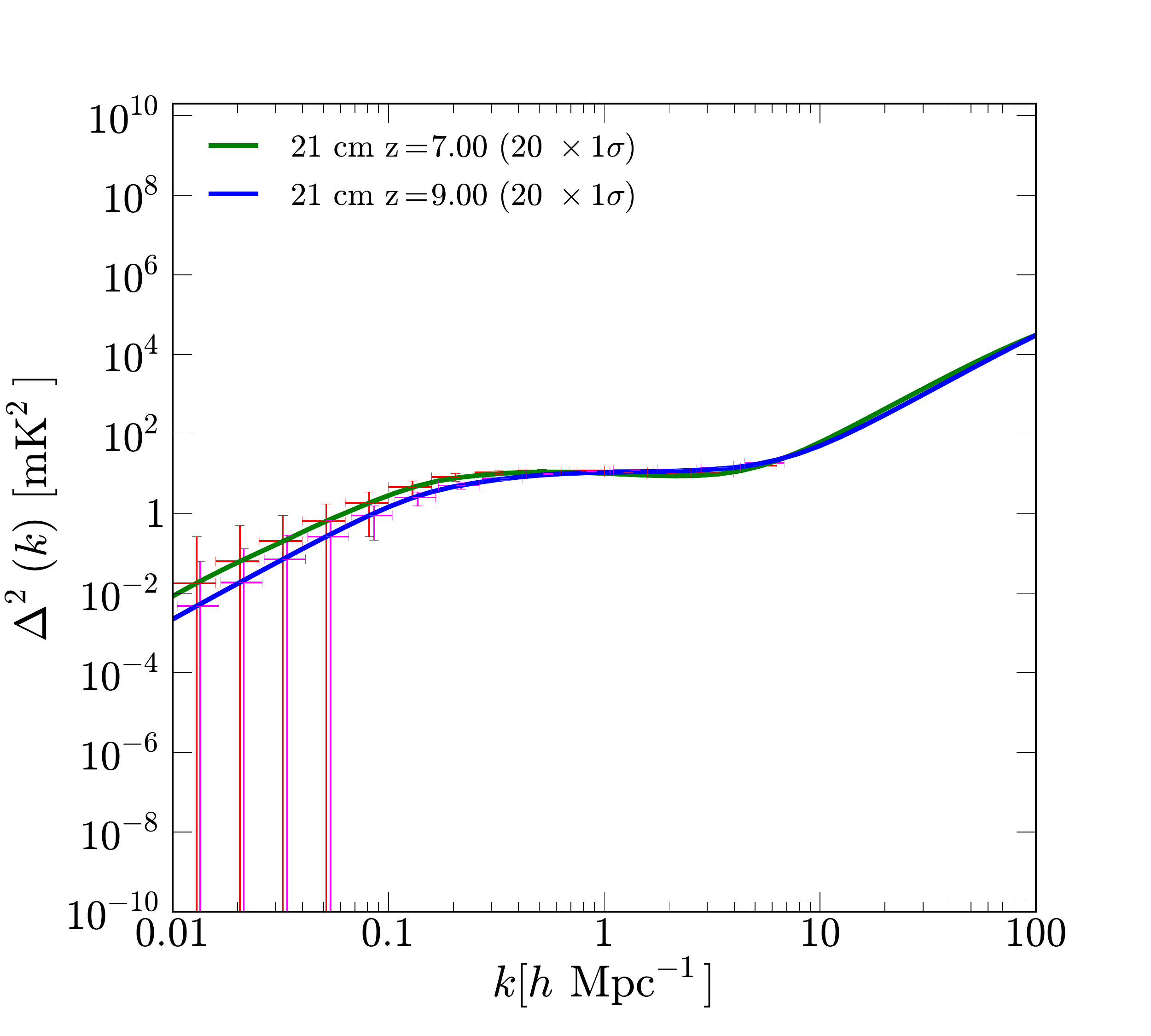}
\caption{Forecast $\HI$ power spectra at $z$ = 7 and 9 for experiment SKA. The multiplicative numbers in parentheses enlarge the error bars for visualization purposes.}\label{sn21cm}
\end{figure}

\begin{figure}
\includegraphics[width=8cm, height=7cm]{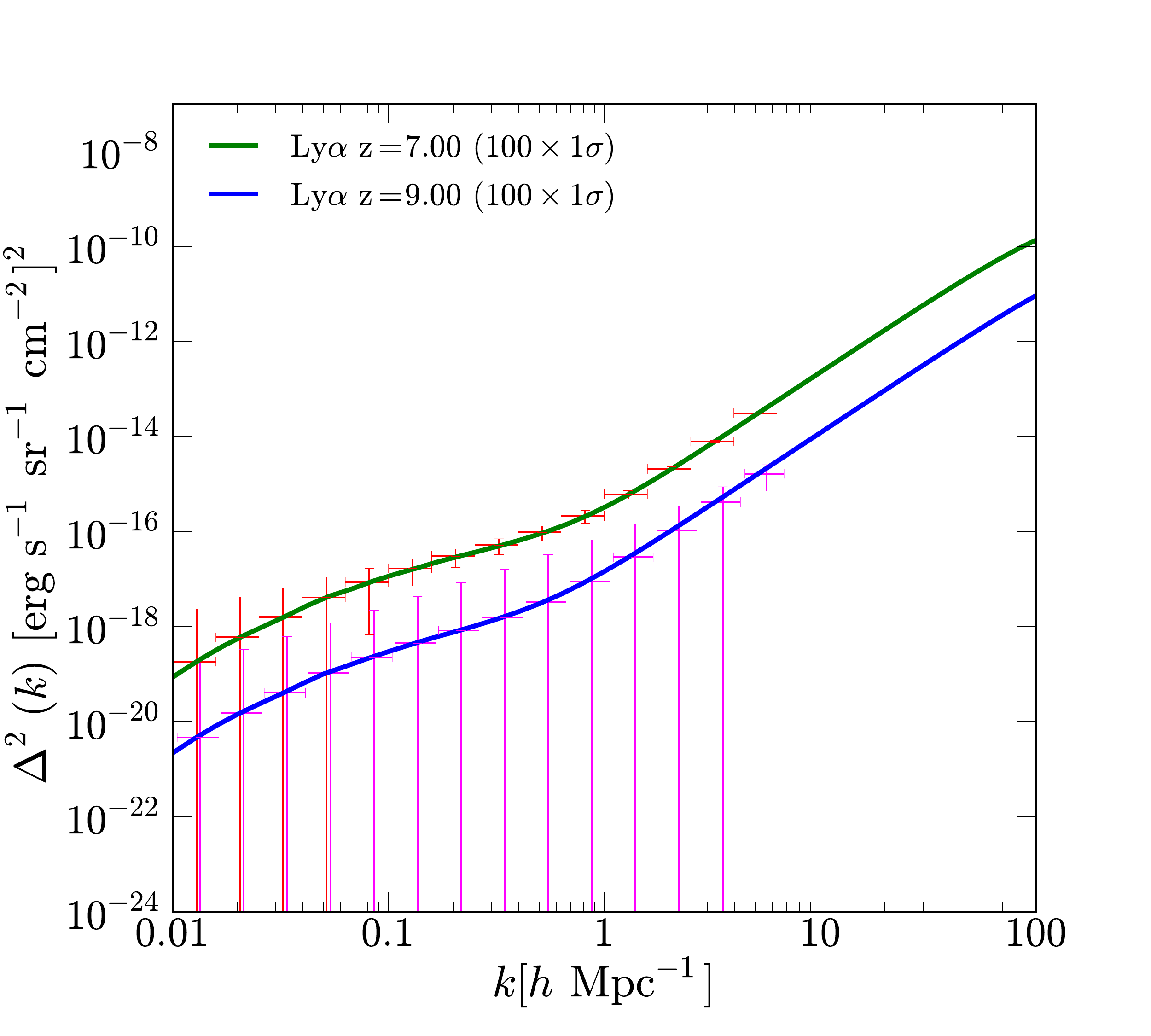}
\caption{Forecast $\ly$ power spectra at $z$ = 7 and 9 for experiment CDIM. The multiplicative numbers in parentheses enlarge the error bars for visualization purposes.}\label{snlya}
\end{figure}

\begin{figure}
\includegraphics[width=8cm, height=7cm]{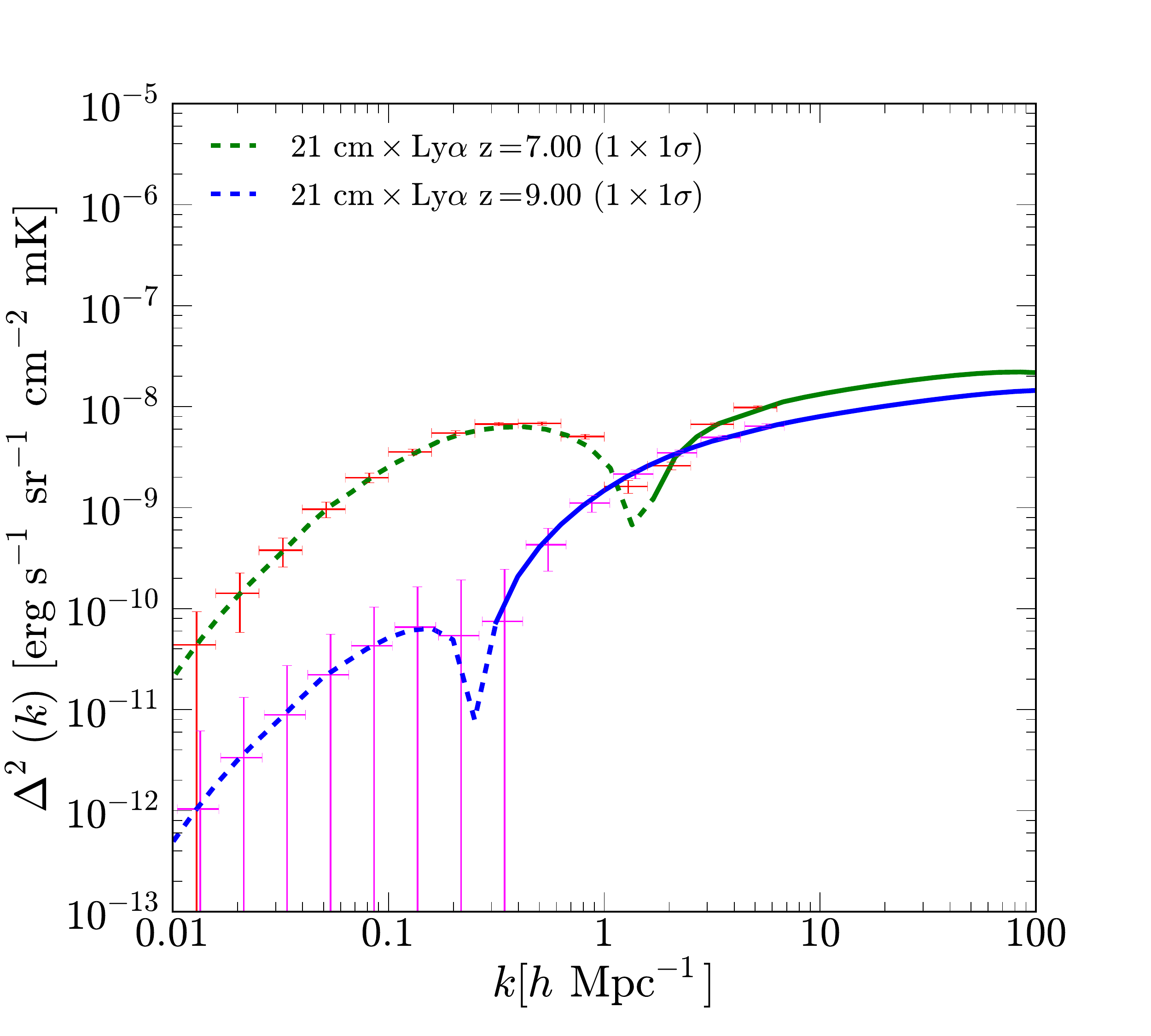}
\caption{Forecast $\HI$\mbox{--}$\ly$ cross-power spectra at $z$ = 7 and 9 for experiments SKA and CDIM. The multiplicative numbers in parentheses enlarge the error bars for visualization purposes.}\label{sncross}
\end{figure}

\section{Forecast for the Experiments}
\label{forecast}

\begin{table*}
\caption{$\HI$ Experiment}\label{t1}
\begin{tabular}{ccccccccccc}
\hline    
 Exp.&$\lambda^{\rm{rest}}\,(\rm{m})$&$D_{\rm{max}}\,(\rm{m})$&$T_{\rm{sys}}\,(\rm{K})$&$t_0\,(\rm{hr})$&$A_{\rm{tot}}\,(\rm{m}^2)$&FOV\,($\rm{deg}^2$)&BW&$\delta\nu$&$N_a$&$z$\\
\hline  
SKA1-LOW&0.21&1000&400&1000&925&13&18MHz&3.9kHz&433&8\\
\hline
\end{tabular}
\end{table*}

\begin{table*}
\caption{$\ly$ Experiment}\label{t2}
\begin{tabular}{ccccccc}
\hline    
 Exp.&$\lambda^{\rm rest}$ (\AA)&$\Omega_{\rm{pix}}\,('')$&$A_s\,(\rm{deg}^2)$&$\sigma_{\rm{pix}}\,(\rm{erg}\,\rm{s}^{-1}\,\rm{cm}^{-2}\,\rm{sr}^{-1})$&$\lambda/\delta{\lambda}$&BW\,($\mu$m)\\
\hline
CDIM&$1216$&1&300&$3\times10^{-6}$&500&0.7-8.0\\
\hline
\end{tabular}
\end{table*}

In this section, we consider two experiments, SKA and Cosmic Dawn Intensity Mapper (CDIM)~\cite{SKA,CDIM}, and investigate the detectability of the $\HI$\mbox{--}$\ly$ cross-correlation with both instrumental noises and foregrounds. We list the experimental specifications in Tables \ref{t1} and \ref{t2}.

The instrumental noise of the $\HI$ experiment is entirely determined by some key factors such as integration time $t_0$, system temperature $T_{\rm{sys}}$, maximum baseline $D_{\rm{max}}$, collecting area $A_{\rm{tot}}$, antenna number $N_a$, and frequency resolution $d\nu$. The noise is then given by
\begin{equation}
P^{\HI}_N=\chi^2y\pi\Big(\frac{\lambda D_{\rm{max}}T_{\rm{sys}}}{A_{\rm{tot}}N_a}\Big)^2\frac{1}{t_0}.
\end{equation}
For the $\ly$ experiments, the noise is
\begin{equation}
P^{\ly}_N=V_{\rm{pix}}\sigma_N^2,
\end{equation} 
where the comoving volume subtended by the detector pixel 
\begin{equation}
V_{\rm{pix}}=\chi^2A_{\rm{pix}}y\delta_{\nu}
\end{equation}
depends on the pixel area $A_{\rm{pix}}$ and frequency resolution $\delta_{\nu}$. The general Knox formula~\cite{2006ApJ...638...20B} for the measured signal $\HI$ or $\ly$ is
\begin{eqnarray}
\Delta P_{X,Y}(k,z)&=&\sqrt{\frac{\tilde P^2_{X,Y}(k,z)+\tilde P_{XX}(k,z)\tilde P_{YY}(k,z)}{N_m}}\label{knox},\nonumber\\ 
\end{eqnarray}
and the number of modes in the bin $k$ is $N_m\,=\,2\pi k^2\Delta k\frac{V_s}{(2\pi)^3}$. Here $X,\,Y$ = $\{\HI,\,\ly\}$, the survey volume is $V_s\,=\,\chi^2A_syB_{\nu}$, $A_s$ is the survey area, and $B_{\nu}$ is the bandwidth (BW). For $\ly$ experiments, the minimum and maximum scales are determined by the survey and pixel areas. For $\HI$, we normally consider modes at scales below $k=10\,\rm{Mpc}^{-1}$ and get the minimum $k$ from the total survey area. For the cross-correlation, the common $k$ range is chosen from two experiments and the minimum volume between the $\HI$ and $\ly$ experiments is taken to calculate the number of modes. The noise- and foreground-included power spectra are formed as $\tilde P_{\HI}\,=\,P_{\HI}+P^N_{\HI}+P^{\rm{radio}}_{\HI}$, $\tilde P_{\ly}\,=\,P_{\ly}+P^N_{\ly}+P^{\rm{low}\hbox{-}z}_{\ly}$, and $\tilde P_{\HI\mbox{--}\ly}\,=\,P_{\HI\mbox{--}\ly}+P^{\rm{radio}\mbox{--}\rm{low}\hbox{-}z}_{\HI\mbox{--}\ly}$. The $k$ region is binned in log-space and we calculate the errors using Eq. (\ref{knox}) for each $k$-band. 

Following Ref.~\cite{2014PhRvD..89b3002D}, we also consider the foreground wedge in Fourier space and its impact on the power spectrum sensitivity. For SKA, the characteristic angle $\theta_0$ is chosen to be $\sqrt{\rm{FOV}}\sim4^{\circ}$. The wedge cut reduces the effective modes $N_m$ in Fourier space and decreases the overall signal-to-noise by 8\%, which could be much larger if a bigger angle $\theta_0$ is assumed. In Figure~\ref{wedge}, we show the reduced region by the foreground wedge using a 2D $\HI$ power spectrum. In addition to the modes cut by the wedge, low $k_{\parallel}$ modes are strongly contaminated by foreground and should be excluded as well. A large $k_{\rm max}$ can compensate for the loss of modes due to the horizontal cut. For example, at $k_{\rm max}\,=\,10\,h\rm{Mpc}^{-1}$,  a horizontal cut between $0.05\,h{\rm Mpc}^{-1}<k_{\parallel}<0.1\,h{\rm Mpc}^{-1}$ introduces negligible changes to the overall signal-to-noise. But for a small $k_{\rm max}$, such as $0.5\,h\rm{Mpc}^{-1}$, the total wedge could reduce the signal-to-noise by 11\% with a horizontal cut at $k_{\parallel}$ = 0.1 $h\rm{Mpc}^{-1}$.

In Figures \ref{sn21cm}, \ref{snlya} and \ref{sncross}, we show all of the power spectra and their band errors for $\HI$ and $\ly$ at $z$ = 7 and 9. As can be seen from Figure~\ref{sncross}, the anti-correlations between neutral hydrogen and galaxies can be probed at very high signal-to-noise ratios.

\begin{figure*}
\includegraphics[width=16cm, height=12cm]{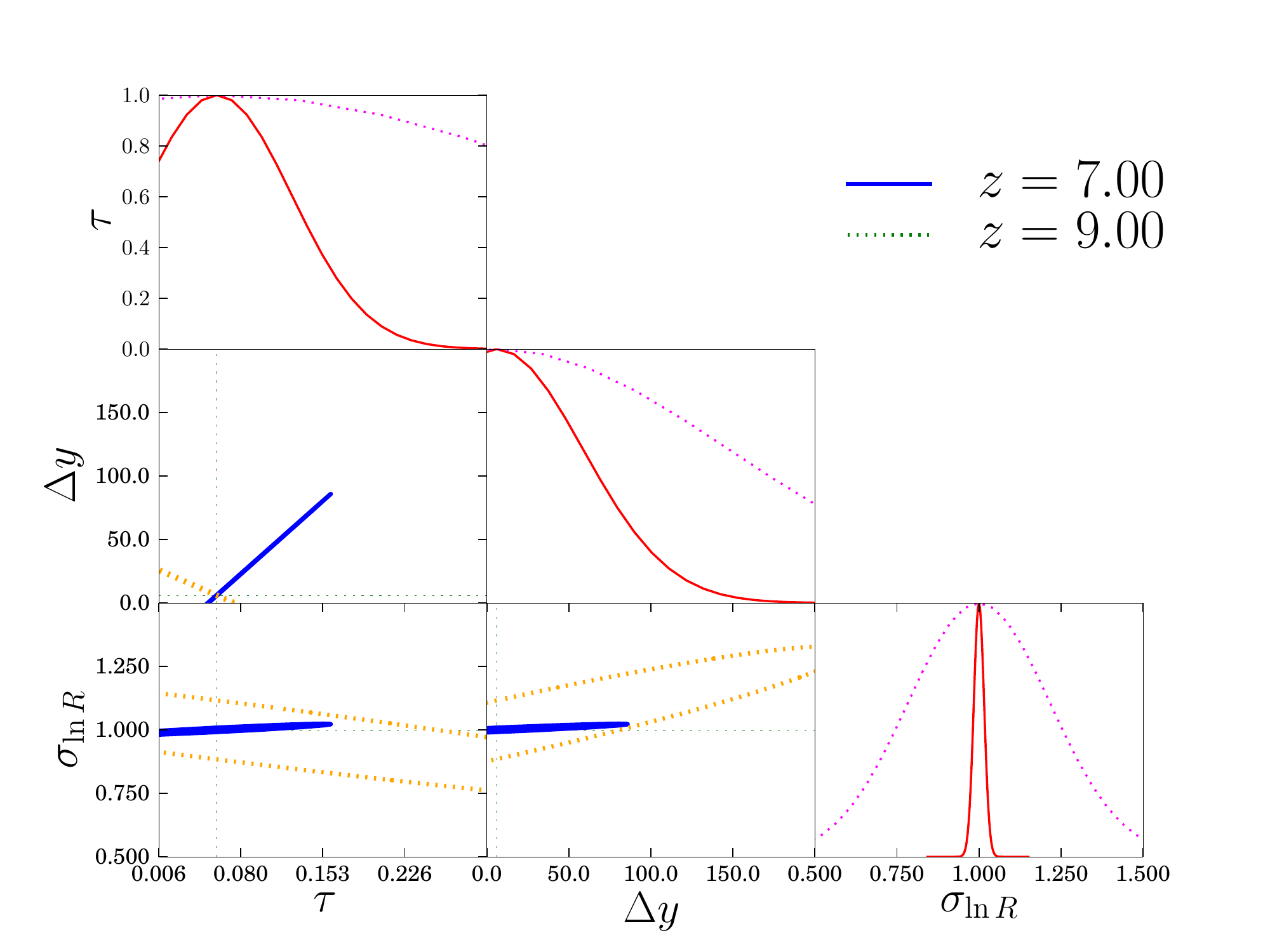}
\caption{1-$\sigma$ confidence levels for the optical depth $\tau$, the EoR duration $\Delta y$, and the rms of the bubble size $\sigma_{\ln R}$ at $z$ = 7 (solid, $\bar x_{\rm{H}}$ = 0.80) and 9 (dotted, $\bar x_{\rm{H}}$ = 0.16).}\label{1s7}
\end{figure*}

\begin{figure*}
\includegraphics[width=16cm, height=12cm]{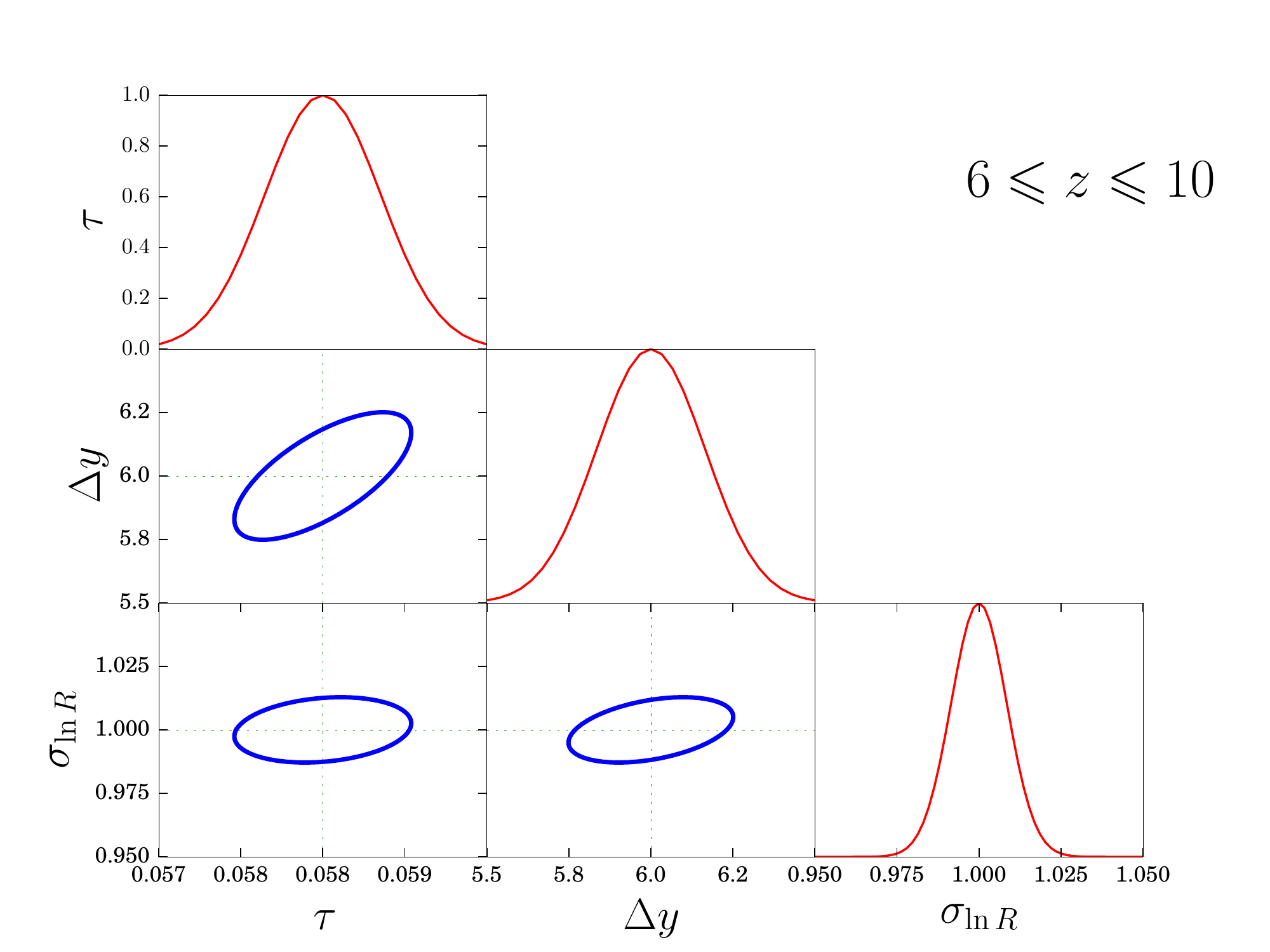}
\caption{1-$\sigma$ confidence levels for the optical depth $\tau$, the EoR duration $\Delta y$, and the rms of the bubble size $\sigma_{\ln R}$ at $6\leqslant z\leqslant 10$.}\label{allz1s7}
\end{figure*}

From the forecasted power spectra in Figure~\ref{sncross}, we can further try to constrain the EoR parameters defined as \textbf{P} = \{$\tau$, $\Delta y$, $\sigma_{\ln R}$\} = \{0.058, 6.0, 1.0\}. The Fisher matrix~\cite{fisher} is 
\begin{equation}
F_{ij}=\displaystyle\sum_{k,z}\frac{1}{(\Delta \eta(k,z))^2}\frac{\partial \Delta^2(k,z)}{\partial p_i}\frac{\partial \Delta^2(k,z)}{\partial p_j}.
\end{equation}
Here $\Delta \eta(k,z)$ is the error on the cross-power spectrum $\Delta^2$ and $p_i$ refers to any parameters in the set \textbf{P} and $\Delta^2$ = $k^3/(2\pi^2)P_{\HI\mbox{--}\ly}$. All of the 1$\sigma$ confidence levels, as well as the likelihood functions in Figure \ref{1s7}, are calculated from the $\HI$\mbox{--}$\ly$ cross-power spectra at $z$ = 7 and 9. As can be seen in the figure, the bubble size can be constrained from the cross-correlation while the errors on the optical depth and duration of the reionization transition are large. This is due to the fact that the cross-correlation is proportional to $x_e(z)$ and not $x^2_e(z)$, to which the $\HI$\mbox{--}$\HI$ and $\HI$\mbox{--}$\tau$~\cite{cora2} are proportional. Therefore, the cross-power spectrum at a single redshift is less sensitive to the reionization history. However, the cross-correlations at different redshifts might be able to break the degeneracies among the parameters, and so are useful as complementary probes to cosmological and astrophysical problems. In Figure~\ref{allz1s7}, we reduce the bandwidth by a factor of 5 and combine the cross-power spectra measured at $z$ = 6, 7, 8, 9, and 10. It is seen that the Fisher matrix error bars on all of the parameters are significantly reduced, and that multi-redshift measurements within narrow bins can effectively break the parameter degeneracies.\\\\

\section{Conclusion}
\label{con}
In this work, we applied a bubble model to the computation of $\HI$ and $\ly$ cross-correlation at the EoR. Making use of the empirical relation between $\ly$ luminosity and mass for the line emissions, we also calculated the power spectra for $\ly$.  The $\HI$\mbox{--}$\ly$ cross-power spectrum in this fast approach can reproduce the key features of the one made by detailed numerical simulations, and we can use it to quickly assess the overall performance of the future EoR experiments. 

The cross-correlation is contaminated by the low-$z$ foregrounds for both $\HI$ and $\ly$. We studied the radio galaxies for the $\HI$ experiments and H$\alpha$, OIII, and OII line emissions for the $\ly$ experiments. All of these foregrounds could be a few orders of magnitude higher than the signals we are probing if the foreground mitigation is not incorporated. The map-space spectral fitting can effectively remove the radio point-source contaminations, while a flux masking for the intensity mapping experiments have been shown to be a good and easy foreground-removal method. 

We take advantage of this efficient algorithm and estimate the errors on the EoR parameters $\tau$, $\Delta y$ and $\sigma_{\ln R}$, based on the Fisher matrix formalism. For other physical processes during the EoR, such as X-ray heating, supernovae explosion, and shock heating, numerical simulations with these effects or an extension to this work should be devised. We will discuss these in the future works.  
\bibliography{lya21}

\section{Acknowledgements}
A.C. and C.F. acknowledge support from NASA grants NASA NNX16AJ69G, NASA NNX16AF39G and Ax Foundation for Cosmology at UC San Diego.

\end{document}